\documentclass[3p,authoryear,times]{elsarticle}
\usepackage{bbm}
\usepackage[utf8]{inputenc}
\usepackage[T1]{fontenc}
\usepackage[hyphens]{url}
\usepackage{tabularx}
\usepackage{tablefootnote}
\usepackage{pdflscape}
\usepackage{booktabs}
\usepackage[svgnames,dvipsnames]{xcolor}
\usepackage[colorlinks]{hyperref}
\colorlet{scolor}{black}
\colorlet{hscolor}{DarkSlateGrey}
\hypersetup{%
  linkcolor={hscolor},
  urlcolor={hscolor},
  citecolor={hscolor},
  filecolor={hscolor},
  menucolor={hscolor},
}
\usepackage{import}
\usepackage{subcaption}
\usepackage{enumitem}
\begin{document}
\begin{frontmatter}
    \title{Bus Stop Spacing Statistics: Theory and Evidence}
    \author[add1]{Saipraneeth Devunuri}
    \author[add1]{Lewis J. Lehe\corref{cor1}}
    \cortext[cor1]{Corresponding author: Lewis J. Lehe, lehe@illinois.edu}
    \author[add1]{Shirin Qiam}
    \author[add1]{Ayush Pandey}
    \author[add2]{Dana Monzer}
    \address[add1]{Department of Civil and Environmental Engineering, University of Illinois Urbana-Champaign}
    \address[add2]{Northwestern University Transportation Center}

    \begin{abstract}
        Discussions of bus stop consolidation sometimes refer to average stop spacings, but there are no reliable statistics about spacings, nor methodologies for calculating them. This paper aims to clarify discussions of bus stop spacings with clear definitions, a methodology for creating statistics from GTFS files and new evidence from real-world systems. It gives national statistics using GTFS data from 539 US transit providers and 83 Canadian providers, as well as particular statistics for thirty providers in the US, ten in Canada and a sample of 38 providers from other countries. US and Canadian spacings are similar. US spacings are wider than sometimes claimed but narrower than those in other countries. Finally, the paper gives examples of metrics created by combining GTFS with data from other sources, and proposes research ideas involving fine-grained stop spacing data.

    \end{abstract}
\end{frontmatter}

\section{Introduction}\label{sec:intro}

``Bus stop spacing'' refers to the distance between consecutive stops on a bus route. The choice of bus stop spacing involves a trade-off: wider spacings (fewer stops) save passengers in-vehicle time (time spent traveling on the bus) but raise passengers access time (time spent traveling to/from stops). This trade-off is at the heart of a robust  academic literature\footnote{See \citet[Sec. 2]{tirachiniEconomicsEngineeringBus2014} for a thorough review.} about the choice of spacings that has been active for about fifty years \citep[e.g.,][]{vuchicRapidTransitInterstation1968,mohringOptimizationScaleEconomies1972, wirasingheSpacingBusStopsMany1981,liAssessingModelOptimal2009,ouyangContinuumApproximationApproach2014,Wu2022}.

In the United States, there is a general perception that current spacings err too far on the side of accessibility at the expense of speed; i.e., that spacings are too narrow. Thus, over the past two decades, many US bus providers have widened their spacings via \emph{stop consolidation}: the systematic practice of removing large numbers of stops. Stop consolidation campaigns have been carried out in Portland \citep{el-geneidyEffectsBusStop2006}, San Francisco \citep{gordonMuniMayReduce2010}, Cincinnati \citep{laFleur2019}, Seattle \citep{kingcountymetroTransitSpeedReliability2021}, Pittsburgh \citep{blazinaPortAuthorityInitial2020}, Dallas \citep{garnhamFightHugeDrop2020}, Denver \citep{RTD}, the Bronx \citep{moloneyFinalBronxBus2021} and elsewhere. Researchers have supported the consolidation trend with various practical techniques for deciding which stops to remove \citep{liAssessingModelOptimal2009,wagnerBenefitcostEvaluationMethod2014, stewartDonStopJust2016}. The stakes in this enterprise have risen over the last decade: buses have lost a non-trivial number of riders to ridehailing \citep{graehler2019understanding,grahn2021travelers,erhardt2022has}, which is more expensive but faster than the bus.

In spite of the vigorous theoretical and practical interest in stop spacing, reliable data about stop spacings is scarce. What is the average American stop spacing? Some studies\footnote{For example: \citet[p. 1730]{furth2000}, \citet[p. 33]{el-geneidyEffectsBusStop2006}, \citet[p. 800]{morencyWalkingTransitUnexpected2011}, \citet[p. 2]{Wu2022}.} reference \citet[p. 4]{reillyTransitServiceDesign1997}: ``It is common European practice to have stops spaced at 3 or 4  per mile [~540-400 m/stop] in contrast with 7 to 10 stops per mile [~230-160 m/stop], which is common in the United States. European bus-stop distances are comparable to rail rapid transit stop distances in the United States.'' Likewise, a Transit Cooperative Research Program (TCRP) report states: ``U.S. transit bus operators...place stops about every 200 m, creating five stops per kilometer [8 stops per mile]'' \citep[p. 4]{NAP10110}. But read in context, both quotes are intended more as impressions than exact statistics, and neither quote references any data collection.

Statistics about real spacings can inform decision-making and communication. \citet{vannesImportanceObjectivesUrban2000} use an analytical model to optimize spacings toward various objectives, then compare the results to their own calculations of average spacings for seven European cities. The real spacings are much shorter than what the model recommends, which gives the theory a practical lesson: that cities ought to consider wider spacings. The \emph{Queens Bus Network Redesign Draft Plan} \citep[p. 9]{queens2022} uses averages to communicate a point to the public:
\begin{quote}
    New York City has too many bus stops, resulting in shorter distances between stops than most other major cities. With an average of 805 feet [245 m] between stops, buses are often stopping as frequently as every one or two blocks. In Queens, the average is slightly higher at 909 feet [277 m]. Both are shorter than the distance between stops in international peer transit systems around the world, which typically range from 1,000 [305 m] to 1,680 feet [512 m].
\end{quote}
The point made here is that the City's status quo is unusual among world cities, so consolidation would not be a harsh experiment. (It is not said where the statistics about the peer transit systems come from or how the averages for Queens and New York were computed.) The international comparison dovetails with the claims from \citet{reillyTransitServiceDesign1997} and \citet{NAP10110} that spacings in other countries are much wider than US spacings. Communications like this are important because changes to stops can be politically fraught \citep{Flint2014,berezBusStopsHere2015}. The nonprofit TransitCenter has even published a report \citep{miatkowskiBusStopBalancing2019} devoted entirely to public communication about consolidation.

At first blush, the dearth of hard data about stop spacings is surprising in light of the trend toward consolidation. The spread of the General Transit Feed Specification (GTFS) \citep{mchugh2013pioneering} has made public transit data easier to analyze, while various API's and websites (e.g., \citet{MobData2023}) have made it possible to download GTFS data from transit providers around the world. But even with GTFS data, it is challenging to compute spacings. In the first place, one cannot simply compute the Haversine distance between each stop, because a particular spacing is (to the degree it speaks to the access/in-vehicle time tradeoff) supposed to be the \emph{driving distance} (i.e., along the route) between two stops. Thus, to calculate spacings involves some complex calculations to obtain driving distances between stops. Also, many agencies do not publish an optional GTFS file (shapes.txt) giving the paths of routes.

Beyond technical challenges, the calculation of stop spacing statistics invites \emph{conceptual} (or definitional) questions. A single route's average spacing can be calculated by dividing its length by its number of stops, but it is not obvious how to take an average for a network. Many stops are served by overlapping routes, or only at certain times, or on certain days. Express routes skip stops. Some routes have portions served only at certain times. Reasonably, frequency could be considered, too. Suppose a bus system has a ``flyer'' that runs thirty kilometers directly from a downtown stop out to an airport twice per day, and also runs a circulator within the downtown that stops every block with ten-minute headways. Should an ``average spacing'' for this system weight the large distance to the airport the same as it does one of the circulator's spacings? Perhaps not: few resources are tied up in the airport route, and its distance is so large that reducing it would not, at the margin, decrease any passengers' walk times.

\citet{pandeyDistributionsBusStop2021}, a preliminary attempt to address these issues, introduces an approach called ``traversal-weighting'' (explained below) and uses GTFS data from 43 US public bus systems to report summary statistics calculated using the \emph{gtfs2gps} R package \citep{Pereira2022}. This paper extends the work begun in \citet{pandeyDistributionsBusStop2021}. This paper's goal is to facilitate further research and discussion of real stop spacings. To this end, this paper has four contributions\begin{enumerate}[label=(\roman*)]
    \item Clarify conceptual issues around stop spacings via new definitions and weighting schemes for the calculation of means.
    \item Devise a methodology for calculating stop spacings using GTFS files, which is implemented in a new Python package \citep{gtfs_segments}.
    \item Provide new evidence about bus stop spacings---mainly in the United States and Canada but also in select cities elsewhere. This evidence is culled from an open-access databases of spacings from 539 providers in the United States \citep{Devunuri2022} and 83 in Canada \citep{Devunuri2023}.
    \item Demonstrate two  alternative spacing statistics: the average spacing \emph{experienced by a passenger} and the number of traffic signals traversed between two stops.
\end{enumerate}

Note the scope of the paper is limited, essentially, to description and clarification. There is no optimization, nor statistical identification (e.g., whether wider spacings cause higher ridership). One motivation for the paper is to provide methods and data by which scholars or practitioners can conduct further statistical and operations research and in the conclusion (Section \ref{sec:conclusion}) we propose several ideas. Another is the more abstract notion that, apart from optimization and statistical identification, data about a topic that affects so many people's transport experience is inherently valuable and worthwhile to collect accurately---even when its applications are not immediately obvious. 

\section{Definitions}\label{sec:theory}

This section discusses conceptual questions that arise in the calculation of statistics and distributions of stop spacings.

\subsection{Terminology}

\begin{figure}
    \centering
    \includegraphics[width=.47\textwidth]{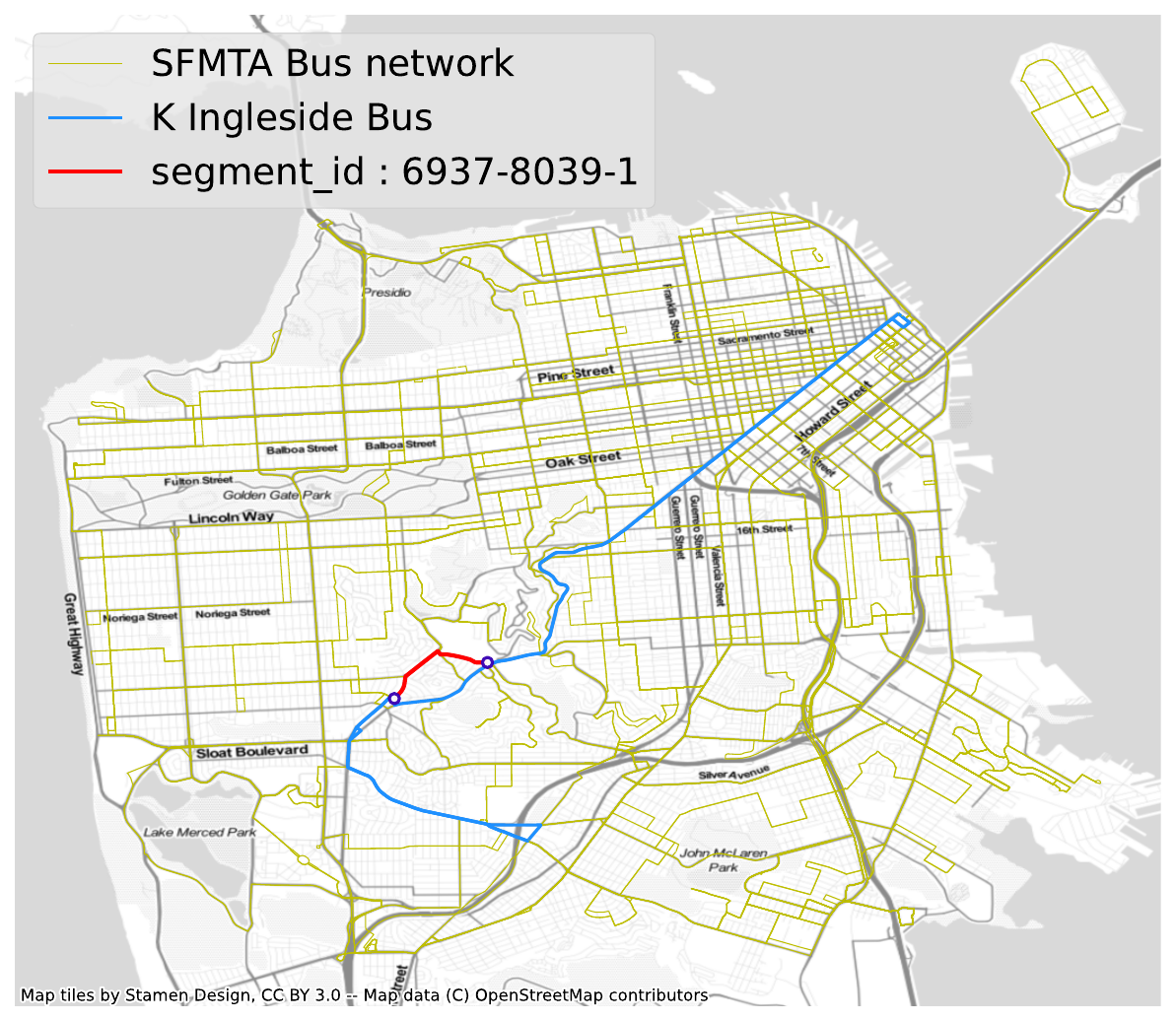}
    \caption{A segment on the K-Ingleside bus route from the San Francisco Municipal Transportation Agency}
    \label{fig:network}
\end{figure}

To begin, we propose the following terminology for the analysis of stop spacings:

\begin{itemize}
    \item \emph{Stop:}  A physical location where buses stop to pick up and drop off passengers.
    \item \emph{Segment:} The edge that a bus travels between visiting two consecutive stops. A segment is characterized by: (i) a ``beginning'' stop (the first stop the bus visits on the segment); (ii) an ``ending'' stop (the second stop the bus visits on the segment); and (iii) a path in space that the bus travels between the two stops. If two routes visit the same two stops consecutively, traveling the same path between them, then we say two routes ``traverse'' the same segment. Figures \ref{fig:network} shows an example segment in red.
    \item \emph{Traversal:} The event of a bus stopping at each of the segment's two stops and traveling along the designated path between them. A segment on a route with a headway of ten minutes (frequency of six buses per hour) will experience twice as many traversals as one on a route with a headway of twenty minutes.
    \item \emph{Distance metric:} The function used to calculate the segment's spacing. Below we focus on \emph{distance along the segment's path}, which takes into account how the path meanders along the road. Section \ref{sec:extension} demonstrates another distance metric: the number of traffic lights traversed by the segment's path.
    \item \emph{Measurement interval}: The part of the schedule during which stop spacing statistics are calculated. Bus schedules vary across days (e.g., Sundays typically have less service than  weekdays) and within days (e.g., some routes and frequencies may only be offered at peaks), so any measurement is specific to a chosen period.
    \item \emph{Deadhead:} Ground covered by a bus when the bus does not carry passengers---when the bus is ``not in service.'' A typical case would be travel to or from a bus depot where there are no pickups or dropoffs. We exclude deadhead mileage from our spacing statistics because they do not reflect the passenger experience.
    \item \emph{Threshold:} In calculating mean spacings, it is sometimes worthwhile to ignore spacings above a certain level which we call the ``threshold.'' The discussion below offers a justification.
\end{itemize}

\subsection{Weighting schemes}

We now use our terminology to describe \emph{weighting schemes} of use in making distributions and summary statistics (e.g., mean, standard deviation, etc.) for a bus system's stop spacings. Suppose that a system has \(N\) segments. Let \(s_i\) be segment \(i\)'s ``spacing'' (most typically, the distance along the bus' path between the two stops) and \(w_i\) be segment \(i\)'s ``weight'' (according to a chosen weighting scheme). In this case, the system's mean spacing is

\begin{equation}\label{eq:No1}
    \bar{s} = \frac{1}{\sum_{i=1}^{N}{w_i}} \cdot \sum_{i=1}^{N}{w_i s_i},
\end{equation}

\noindent and the weighted cumulative distribution of stop spacing \(s\) is
\begin{equation}\label{eq:ecdf}
    F(s) = \frac{1}{\sum_{i=1}^N w_i} \cdot \sum_{i=1}^N w_i \mathbbm{1}_{s_i \leq s},
\end{equation}
where \(\mathbbm{1}_{s_i \leq s}\) is an indicator equal to 1 if \(s_i\leq s\).

Now consider the following three weighting schemes. These are not the only possible weighting schemes, but they are intuitive and can be calculated directly from most GTFS files.

\begin{itemize}
    \item \emph{Segment-weighted}: Each segment receives the same weight: i.e., $w_i=1 \, \forall i$.

    \item \emph{Route-weighted}: Each segment is weighted by the number of routes that include it. The mean route-weighted stop spacing could be calculated by summing up the lengths of all routes (ignoring deadhead portions) and then dividing by the total number of segments. This is the mean distance between stops that someone would experience if they rode every route in the schedule once.

    \item \emph{Traversal-weighted}: Each segment is weighted by the number of times a bus traverses it---i.e., by its total number of traversals in the schedule. Stops on high-frequency routes are thus weighted more heavily. The mean traversal-weighted stop spacing can also be calculated by summing the cumulative distances all buses travel (ignoring deadhead portions) and dividing by the total number of segments. The mean traversal-weighted stop spacing is also the average distance that a bus travels between consecutive stops.

\end{itemize}

\begin{figure}[ht]
    \centering
    \includegraphics[width=0.6\textwidth]{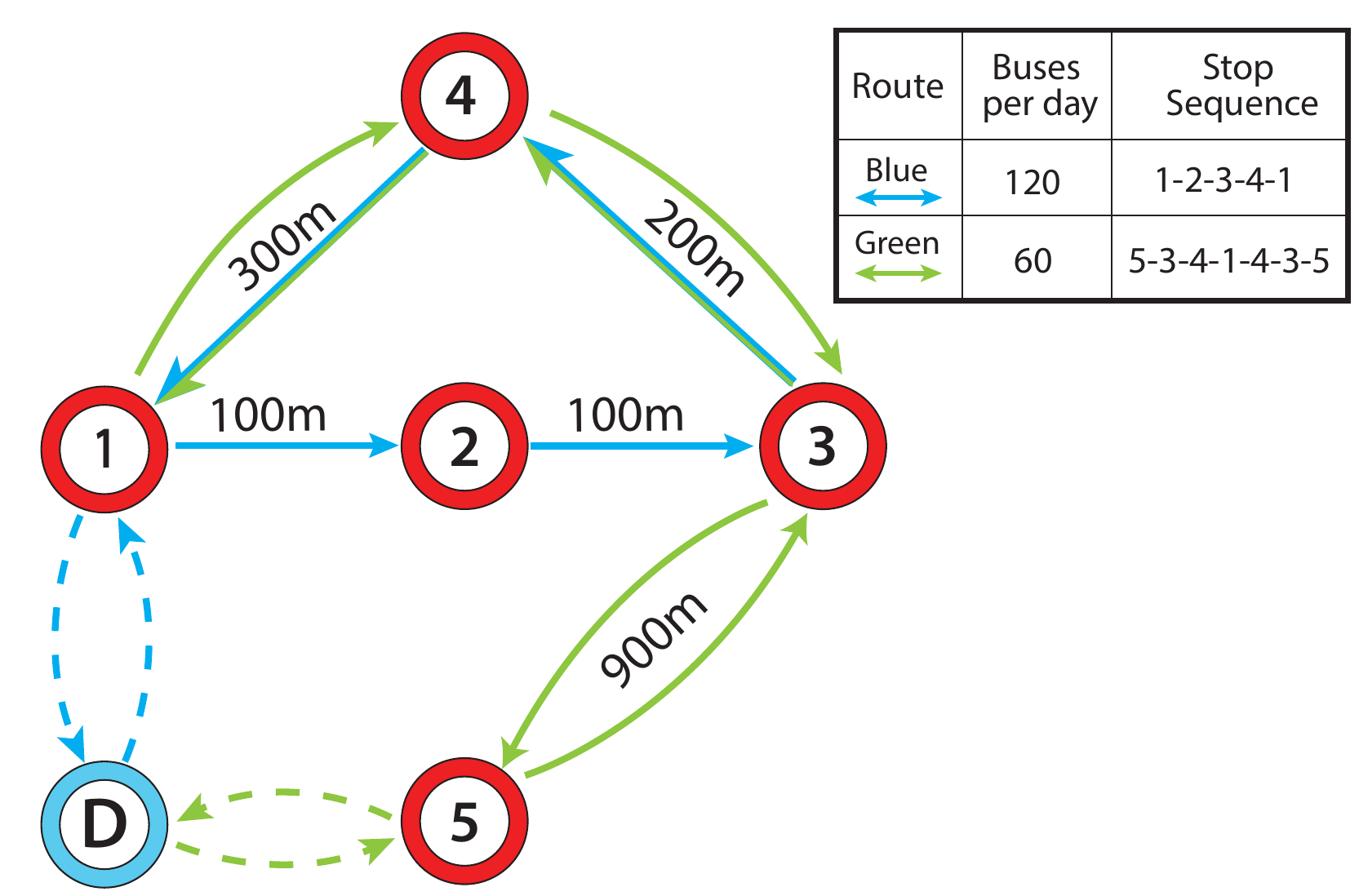}
    \caption{Example network with two bus lines (not drawn to scale)}\label{fig:HypotheticalNetwork}
\end{figure}

Figure \ref{fig:HypotheticalNetwork} shows a simple network to illustrate the weighting schemes. There are two routes: Green and Blue. There are five stops, named by numbers, and one depot named \textbf{D}. The ``measurement interval'' chosen is some hypothetical day (e.g., a weekday), during which Green runs 60 buses and Blue 120 buses per day. There are eight segments---each depicted as an arrow on the map, with colors representing what route serves the segment. Both routes traverse the segments from 3 to 4 and from 4 to 1, so these segments are dual-colored. Each segment's spacing is written alongside its arrow. The figure is not drawn to scale. Dashed arrows to/from the depot indicate deadhead segments. The table beside the figure shows the sequence of stops on each route.

\begin{table}[ht]
    \centering
    \caption{Mean spacings for the example network}\label{tab:hypoNet}
    \small
    \begin{tabular}{clc|ccc|}
\cline{4-6}
\multicolumn{1}{l}{}       &                                      & \multicolumn{1}{l|}{} & \multicolumn{3}{c|}{\textbf{Weight of Segment}}          \\ \hline
\multicolumn{1}{|c|}{\textbf{Segment}} &
  \multicolumn{1}{l|}{\textbf{Route}} &
  \textbf{Spacing {[}m{]}} &
  \multicolumn{1}{c|}{\textbf{Segment-}} &
  \multicolumn{1}{c|}{\textbf{Route-}} &
  \textbf{Traversal-} \\ \hline
\multicolumn{1}{|c|}{1\_2} & \multicolumn{1}{l|}{Blue}          & 100                   & \multicolumn{1}{c|}{1} & \multicolumn{1}{c|}{1} & 120    \\ \hline
\multicolumn{1}{|c|}{2\_3} & \multicolumn{1}{l|}{Blue}          & 100                   & \multicolumn{1}{c|}{1} & \multicolumn{1}{c|}{1} & 120    \\ \hline
\multicolumn{1}{|c|}{3\_4} & \multicolumn{1}{l|}{Blue \& Green} & 200                   & \multicolumn{1}{c|}{1} & \multicolumn{1}{c|}{2} & 120+60 \\ \hline
\multicolumn{1}{|c|}{4\_1} & \multicolumn{1}{l|}{Blue \& Green} & 300                   & \multicolumn{1}{c|}{1} & \multicolumn{1}{c|}{2} & 120+60 \\ \hline
\multicolumn{1}{|c|}{5\_3} & \multicolumn{1}{l|}{Green}           & 900                   & \multicolumn{1}{c|}{1} & \multicolumn{1}{c|}{1} & 60     \\ \hline
\multicolumn{1}{|c|}{4\_3} & \multicolumn{1}{l|}{Green}           & 200                   & \multicolumn{1}{c|}{1} & \multicolumn{1}{c|}{1} & 60     \\ \hline
\multicolumn{1}{|c|}{1\_4} & \multicolumn{1}{l|}{Green}           & 300                   & \multicolumn{1}{c|}{1} & \multicolumn{1}{c|}{1} & 60     \\ \hline
\multicolumn{1}{|c|}{3\_5} & \multicolumn{1}{l|}{Green}           & 900                   & \multicolumn{1}{c|}{1} & \multicolumn{1}{c|}{1} & 60     \\ \hline
\multicolumn{3}{|c|}{\textbf{Average Stop Spacing   (m)}} &
  \multicolumn{1}{c|}{\textbf{375}} &
  \multicolumn{1}{c|}{\textbf{350}} &
  \textbf{300} \\ \hline
\end{tabular}%

\end{table}

Consider next Table \ref{tab:hypoNet}. Segments are named by the scheme \(a\_b\), where \(a\) is the beginning stop and \(b\) is the ending stop. Because segments are directionally-specific, stops such as 1 and 4 are linked by two \emph{different} segments---one of which (4\_1) is traversed more often. The last three columns show each segment's weight according to each of the weighting schemes described above. The bottom row shows the mean spacings as calculated by each weighting scheme. Note the means are substantially different from one another. The traversal-weighted mean is the smallest, because the shorter segments have more traversals. The route-weighted mean is the next largest, and the segment-weighted is the largest.

Which of the three means is the best? Ultimately, there is no ``true,'' ``natural'' or ``genuine'' mean. But among the three, traversal-weighting has the advantage of being tied to a physical event: the traversal-weighted mean is the mean distance that a bus travels between stops. If planners decided to cosmetically divide a route into two routes with different names, then the route-weighted mean spacing would change even if nothing about the passenger experience changed. Also, unlike segment-weighting, the traversal-weighted mean favors the parts of a network that have the most service. So if a provider occasionally runs a bus out to a distant suburb, the traversal-weighted mean is less affected than the segment-weighted is.

\subsection{Threshold}

\begin{figure}[ht]
    \centering
    \includegraphics[width=.8\textwidth]{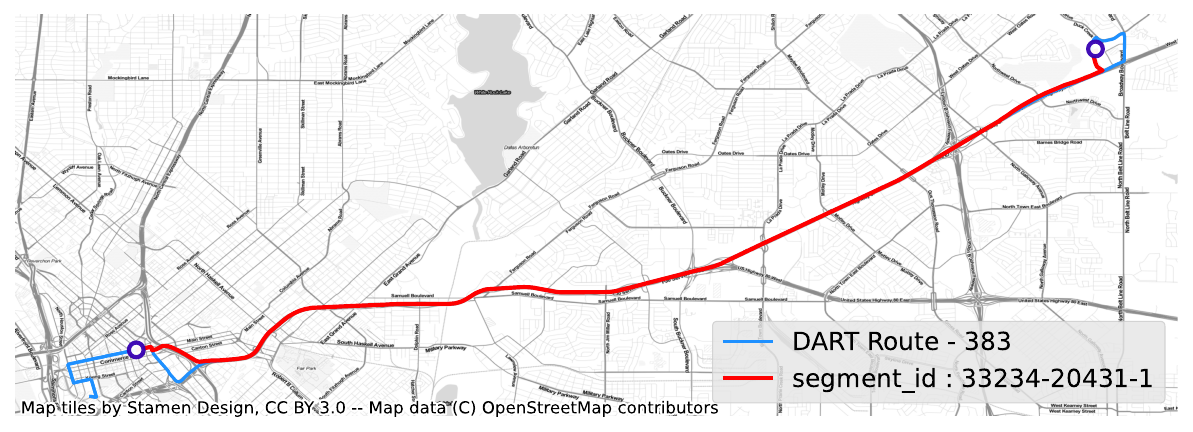}
    \caption{20.9 km segment on DART Express Bus Route 383}\label{fig:dart}
\end{figure}

The last concept we propose is the \emph{threshold}. The threshold is the upper bound of spacing used in calculating a ``truncated'' mean (or other statistic). For example, later, we give a mean spacing of all segments shorter than 2 km for real bus service providers. The benefit of using a truncated mean to describe a network is best captured by an edge case. Figure \ref{fig:dart} shows a segment of a DART express bus route running from a suburban bus station into downtown Dallas. The segment is about 21 km long and runs almost entirely on Interstate 30. In considering stop policies for DART, or in comparing DART to other networks, it may be helpful to use a mean that excludes segments such as this one. In the first place, the segment is an extreme statistical outlier that will skew results. Second, the utility of using stop spacing statistics is largely related to the question of consolidation, and there is no trade-off between time spent walking to/from stops and in-vehicle time for marginal changes in spacing along this segment. Setting aside the fact that no one walks along Interstate 30, the segment is so long that no one would walk to a stop from the middle of the segment, so somewhat increasing or decreasing the spacing would not substantially change anyone's walk times. Of course, the entire route 383 need not be excluded: upon leaving the interstate, the bus traverses a handful of closely-spaced segments; their spacings are both typical for DART and relevant to the trade-off between access and in-vehicle time.

\section{Methodology}\label{sec:methodology}

Following the definitions proposed above, this section describes a methodology for data collection, processing, and analysis to use GTFS files for stop spacing statistics. We have created a python package called \emph{gtfs-segments} which replicates the methodology in this paper to download the latest sources, process data, and calculate stop spacings. The package is publicly hosted on GitHub at \url{https://github.com/UTEL-UIUC/gtfs_segments}. The GitHub repository hosts detailed documentation for readers interested in using the package.


The GTFS data for a transit provider is organized as folder (usually stored as a zip file) which contains various tabular text files created in accordance with the GTFS Schedule\footnote{See \url{https://gtfs.org/schedule/} for more details.} specification. Of these, the files required for \emph{gtfs-segments} to run are stops.txt, routes.txt, trips.txt, stop\_times.txt, calendar.txt (or calendar\_dates.txt), and shapes.txt. The last of these, shapes.txt, is listed as ``optional'' in the GTFS Schedule specification, and some agencies do not include it in their public feeds. It contains trip shapes: latitude/longitude coordinates for the paths of all routes. These shapes are used to calculate the distances between stops and to split routes into segments. Every shape has a unique id that corresponds to a particular trip.

The GTFS feed has schedules for all days of operations between the feed start and end dates. To compare spacings between different cities, we only use trips served during the \emph{busiest day of the schedule}, defined to be the day with the most service\_ids. In every case, the busiest day is a weekday. Using \emph{calendar.txt} and \emph{calendar\_dates.txt}, we identify the busiest day in the schedule and obtain the corresponding service\_ids. These service\_ids are used to filter the trips in \emph{trips.txt}, and segments for each trip are created based on consecutive stops in `stop\_times.txt.' The `stop\_times.txt' file includes optional columns called `pickup\_type' and `dropoff\_type', which indicate trips and scheduled stops for which service is not provided. We use these to try and eliminate deadhead segments. Trips with only one stop time listed are also eliminated as non-revenue trips.

\begin{figure}[h]
    \centering
    \includegraphics[width=0.6\textwidth]{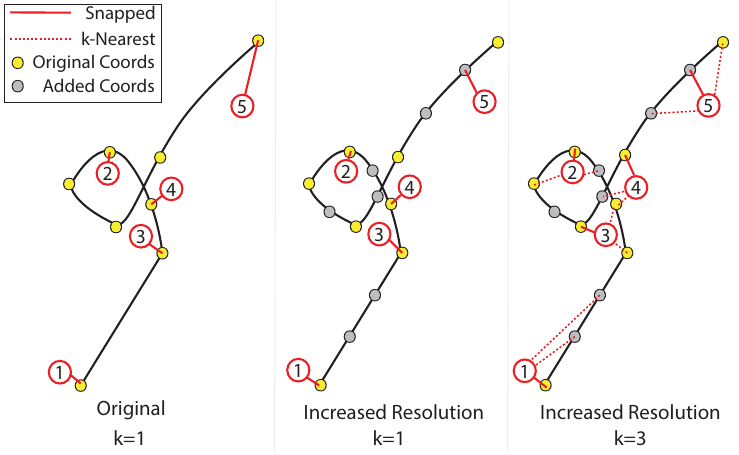}
    \caption{Improvement in snapping due to increase in resolution and k-nearest neighbors}\label{fig:snap}
\end{figure}

With the data downloaded and relevant trips identified, we proceed to split trip shapes into `segments' (as defined in Section \ref{sec:theory}) which can be measured. This splitting is challenging because GTFS does not pair the stop locations (latitude/longitude pairs) in the stops.txt file to points on each trip's shape. Stops thus need to be ``snapped'' onto the trip's shape before segmentation. Intuitively, one could project each stop's location onto the nearest point on the trip shape or match each stop to the shape's nearest coordinate. However, both approaches may lead to errors whereby a stop is mapped to an out-of-order point on the trip shape. This happens especially when trips involve loops.

Figure \ref{fig:snap} shows an example route with five stops in the sequence $1\rightarrow 5$. The coordinates of a the trip shape are represented as yellow dots, and the coordinates of stops are represented as numbered dots. In the first panel, snapping each stop to the nearest given coordinate leads to two problems: First, the stop ordering is incorrect, because the bus visits stops 3 and 4 before stop 2. Second, stop 5 is snapped to a point which is far off from its projection onto the trip shape. To overcome these challenges, we use the following procedure.

\begin{itemize}
    \item First, we increase the resolution of the trip shape by adding points (via linear interpolation between the given coordinates) until no two consecutive points are separated by more than \emph{five} meters. Added points are indexed appropriately to indicate the ordering of points along the route. In the second panel of Figure \ref{fig:snap}, the added points are represented by grey dots. Increasing the resolution successfully snaps 5 to a closer point along the route, but it does not solve the problem of out-of-order stops.
    \item Next, to order stops correctly, we transform the shape's coordinates (including the added coordinates) into a \emph{k-dimensional tree data structure}\footnote{ We used the `spatial.KDTree' function from the scipy package}. As a first pass, we use \(k=3\). Using this tree, the k-nearest neighbors (i.e., the three closest points on the shape) are identified for each stop location in the trip. Among the \(k\) neighbors, we choose the closest one to the stop (using Haversine distance) which is also in the correct order. If, for any stop, there are no neighbors on the trip shape in the correct order, then we double the \(k\) value and start over. This process continues either until we have a suitable ordering or we discard the trip if no solution is found even with \(k\) set equal to the number of coordinates given for the entire route. In practice, for a typical agency only a handful of trips are discarded, and upon manual inspection we have noticed these are almost always mislabeled somehow (i.e., cutting through lots). Notice in the last panel of Figure \ref{fig:snap}, the increase in resolution and using k-nearest neighbors with \(k=3\) snaps the stops in the correct order.
    \item Finally, once every stop has been snapped to a coordinate on the trip shape, the intermediate points between stops are collected into segments. These are stored as LineStrings that represent the geographic path of the segment. All LineStrings are projected onto the Mercator projection (EPSG:4326/WGS 84) to ensure consistency. The distance metric of each LineString is then calculated and recorded as the segment's spacing.
\end{itemize}

The process above takes about three to five minutes on a standard desktop PC for a large transit system such as the Chicago Transit Authority. When processing is finished, we produce a table in the form of a GeoPandas GeoDataFrame \citep{kelsey_jordahl_2020_3946761}. The GeoDataFrame is a Python object with methods for filtering, sorting, grouping, and visualizing geospatial data. We use this to compute statistics.

\section{Results }\label{sec:results}
Following the methodology detailed above, we have created open-source datasets containing all segments in the United States and Canada and hosted them on the Harvard Dataverse service. The US repository is hosted at \citet{Devunuri2022}, and the Canadian one at \citet{Devunuri2023}. All feeds used were updated in late 2022 or early 2023 and not listed as ``deprecated'' on the MobilityData Database \citep{MobData2023}.  Each provider's data is gleaned from GTFS feeds updated in late 2022 or 2023. In the dataset, each provider has one folder. Each folder contains both tabular and GeoJSON files for examining and mapping the provider's data, histograms of spacings, and summary statistics calculated using each of the weighting schemes described above. In total, our data consisted of 660 feeds from 539 providers in the US, 83 in Canada, and 38 in cities in other countries.


National means and medians, calculated using our three weighting systems, appear in Table \ref{tab:sum_whole} for the 539 US and 83 Canadian bus providers in our database. The ``threshold'' chosen for calculating these statistics is 2 km: all segments longer than 2 km are excluded. We chose this threshold because it excludes fewer than 2\% of segments in nearly every city (thereby giving a complete view) but still excludes segments too long for most people to walk from halfway between the stops (i.e., segments for which the marginal change in spacing would not result in shorter realized walk times).


Note the national statistics reported in Table \ref{tab:sum_whole} are weighted by their corresponding measures: i.e., the traversal-weighted mean is \emph{not} the mean traversal-weighted mean of all cities, but rather the mean spacing traversed in the US overall. The three means and medians for both countries are ordered as they were in the example network above: the traversal-weighted means are narrower than the route-weighted means, which are narrower than the segment-weighted means. This ordering is intuitive: relative to suburban routes, segments located in dense urban areas are shorter, traversed by more routes, and traversed more frequently. Table \ref{tab:units} converts between units which may be more intuitive. 

\begin{table}[h]
    \centering
    \small
    \caption{Summary statistics for the United States (Threshold 2 km)}\label{tab:sum_whole}
    \begin{tabular}{|l|c|c|}
    \hline
    {\textbf{Attribute}} & \multicolumn{2}{c|}{\textbf{Value}}                \\ \cline{2-3} 
                                    & \multicolumn{1}{c|}{\textbf{US}} & \textbf{Canada} \\ \hline

    traversal-weighted mean {[}m{]}                  & 352 & 347           \\ \hline
    route-weighted mean {[}m{]}                      & 389 & 353         \\ \hline
    segment-weighted mean {[}m{]}                    & 401 & 367            \\ \hline

    traversal-weighted median {[}m{]}                  & 295 & 287           \\ \hline
    route-weighted median {[}m{]}                      & 317 & 290         \\ \hline
    segment-weighted median {[}m{]}                    & 326 & 298            \\ \hline
    
    \% segments w/ spacing \textgreater threshold (2 km) & 1.89  & 1.50   \\ \hline
\end{tabular}%
\end{table}
\begin{table}[h]
    \centering
    \small
    \caption{Unit conversions}\label{tab:units}
    \begin{tabular}{|l|c|c|c|c|c|c|c|c|c|}
        \hline
        \textbf{stops/mile} & \textbf{10} & \textbf{9} & \textbf{8} & \textbf{7} & \textbf{6} & \textbf{5} & \textbf{4} & \textbf{3} & \textbf{2} \\ \hline
        meters/stop         & 161         & 179        & 201        & 230        & 268        & 322        & 402        & 536        & 805        \\ \hline
        feet/stop           & 528         & 587        & 660        & 754        & 880        & 1,056      & 1,320      & 1,760      & 2,640      \\ \hline
        stops/km            & 6.2         & 5.6        & 5.0        & 4.3        & 3.7        & 3.1        & 2.5        & 1.9        & 1.2        \\ \hline
    \end{tabular}
\end{table}

Table \ref{tab:US_cities} lists summary statistics spacings for thirty of the largest bus service providers by ridership in the United States, and Table \ref{tab:CA_cities} does the same for the ten largest providers in Canada.  The cities are listed in increasing order of their traversal-weighted mean spacing. Note that, to save space, we have left route-weighted statistics out of Tables \ref{tab:US_cities} and \ref{tab:CA_cities}, because---as noted above---this system is more arbitrary than the other two, being dependent on how routes are named\footnote{One way to ameliorate this arbitrariness is to group trips by their `route\_short\_name' property, but many agencies do not provide this property.} in the GTFS files.

\begin{table}[ht]
    \caption{Stop Spacings for top 30 transit providers in the United States (threshold  2km).}\label{tab:US_cities}
    \resizebox{\columnwidth}{!}{%
        \begin{tabular}{|ll|c|c|c|c|c|}
\hline
\multicolumn{1}{|l|}{\textbf{Provider}} &
  \textbf{Urbanized Area} &
  \textbf{\begin{tabular}[c]{@{}c@{}}Traversal \\ Weighted \\ Mean {[}m{]}\end{tabular}} &
  \textbf{\begin{tabular}[c]{@{}c@{}}Segment \\ Weighted \\ Mean {[}m{]}\end{tabular}} &
  \textbf{\begin{tabular}[c]{@{}c@{}}Traversal \\ Weighted \\ Median {[}m{]}\end{tabular}} &
  \textbf{\begin{tabular}[c]{@{}c@{}}Segment \\ Weighted \\ Median {[}m{]}\end{tabular}} &
  \textbf{\begin{tabular}[c]{@{}c@{}}\% Segments \\ w/ spacing \textgreater \\ threshold \\ (2km)\end{tabular}} \\ \hline
\multicolumn{1}{|l|}{Southeastern Pennsylvania Transportation Authority (SEPTA)} &
  Philadelphia, PA &
  217 &
  274 &
  172 &
  193 &
  0.313 \\ \hline
\multicolumn{1}{|l|}{Chicago Transit Authority (CTA)} &
  Chicago, IL &
  230 &
  242 &
  206 &
  206 &
  0.142 \\ \hline
\multicolumn{1}{|l|}{San Francisco Municipal Transportation   Agency (SFMTA)} &
  San Francisco, CA &
  261 &
  264 &
  219 &
  209 &
  0.368 \\ \hline
\multicolumn{1}{|l|}{Pittsburgh Regional Transit (PRT)} &
  Pittsburgh, PA &
  273 &
  304 &
  209 &
  221 &
  1.112 \\ \hline
\multicolumn{1}{|l|}{Massachusetts Bay Transportation   Authority (MBTA)} &
  Boston, MA &
  284 &
  301 &
  239 &
  239 &
  0.461 \\ \hline
\multicolumn{1}{|l|}{\begin{tabular}[c]{@{}l@{}}MTA - New York City Transit (NYCT) \& MTA Bus\\      Company\end{tabular}} &
  New York, NY &
  289 &
  317 &
  236 &
  244 &
  0.62 \\ \hline
\multicolumn{1}{|l|}{Metro Transit} &
  Minneapolis, MN &
  298 &
  294 &
  227 &
  226 &
  0.635 \\ \hline
\multicolumn{1}{|l|}{Washington Metropolitan Area Transit Authority (WMATA)} &
  Washington, DC &
  300 &
  321 &
  245 &
  257 &
  0.416 \\ \hline
\multicolumn{1}{|l|}{Metropolitan Atlanta Rapid Transit Authority (MARTA)} &
  Atlanta, GA &
  308 &
  315 &
  263 &
  265 &
  0.572 \\ \hline
\multicolumn{1}{|l|}{METRO Houston} &
  Houston, TX &
  309 &
  326 &
  261 &
  268 &
  0.758 \\ \hline
\multicolumn{1}{|l|}{Dallas Area Rapid Transit (DART)} &
  Dallas, TX &
  310 &
  319 &
  257 &
  263 &
  0.504 \\ \hline
\multicolumn{1}{|l|}{Bi-State Development Agency (METRO)} &
  St. Louis, MO &
  316 &
  334 &
  264 &
  277 &
  0.368 \\ \hline
\multicolumn{1}{|l|}{Milwaukee County Transit System (MCTS)} &
  Milwaukee, WI &
  326 &
  342 &
  296 &
  303 &
  0.187 \\ \hline
\multicolumn{1}{|l|}{TriMet} &
  Portland, OR &
  335 &
  349 &
  285 &
  294 &
  0.868 \\ \hline
\multicolumn{1}{|l|}{Maryland Transit Administration} &
  Baltimore, MD &
  339 &
  393 &
  276 &
  305 &
  0.765 \\ \hline
\multicolumn{1}{|l|}{Alameda-Contra Costa Transit District (AC Transit)} &
  Oakland, CA &
  348 &
  377 &
  294 &
  307 &
  0.54 \\ \hline
\multicolumn{1}{|l|}{New Jersey Transit (NJ Transit)} &
  Newark, NJ &
  350 &
  448 &
  265 &
  331 &
  2.194 \\ \hline
\multicolumn{1}{|l|}{VIA Metropolitan Transit (VIA)} &
  San Antonio, TX &
  353 &
  349 &
  275 &
  276 &
  1.227 \\ \hline
\multicolumn{1}{|l|}{TheBus} &
  Honolulu, HI &
  359 &
  376 &
  283 &
  286 &
  1.379 \\ \hline
\multicolumn{1}{|l|}{Broward County Transit} &
  Fort Lauderdale, FL &
  362 &
  371 &
  307 &
  309 &
  0.829 \\ \hline
\multicolumn{1}{|l|}{Miami-Dade Transit} &
  Miami, FL &
  372 &
  399 &
  299 &
  319 &
  1.254 \\ \hline
\multicolumn{1}{|l|}{Metro} &
  Los Angeles, CA &
  381 &
  396 &
  352 &
  363 &
  0.926 \\ \hline
\multicolumn{1}{|l|}{Regional Transportation District (RTD)} &
  Denver, CO &
  390 &
  412 &
  346 &
  353 &
  1.147 \\ \hline
\multicolumn{1}{|l|}{King County Metro Transit} &
  Seattle, WA &
  391 &
  406 &
  345 &
  348 &
  0.876 \\ \hline
\multicolumn{1}{|l|}{San Diego Metropolitan Transit System (MTS)} &
  San Diego, CA &
  406 &
  429 &
  341 &
  355 &
  1.763 \\ \hline
\multicolumn{1}{|l|}{Valley Metro (VM)} &
  Phoenix, AZ &
  433 &
  451 &
  401 &
  405 &
  0.449 \\ \hline
\multicolumn{1}{|l|}{Orange County Transportation Authority (OCTA)} &
  Orange County, CA &
  441 &
  451 &
  386 &
  393 &
  0.731 \\ \hline
\multicolumn{1}{|l|}{Regional Transportation Commission of Southern Nevada (RTC)} &
  Las Vegas, NV &
  444 &
  441 &
  362 &
  367 &
  0.24 \\ \hline
\multicolumn{1}{|l|}{Santa Clara Valley Transportation Authority (VTA)} &
  San Jose, CA &
  463 &
  483 &
  395 &
  404 &
  1.331 \\ \hline
\multicolumn{1}{|l|}{Capital Metro} &
  Austin, TX &
  484 &
  506 &
  396 &
  402 &
  1.944 \\ \hline
\multicolumn{2}{|c|}{\textbf{Summary}} &
  \textbf{320} &
  \textbf{355} &
  \textbf{269} &
  \textbf{288} &
  \textbf{0.842} \\ \hline
\end{tabular}%
    }
\end{table}

\begin{table}[ht]
    \caption{Stop Spacings for 10 transit providers in Canada (threshold 2km).}\label{tab:CA_cities}
    \resizebox{\columnwidth}{!}{%
        \begin{tabular}{|ll|c|c|c|c|c|}
\hline
\multicolumn{1}{|l|}{\textbf{Provider}} &
  \textbf{Urbanized Area} &
  \textbf{\begin{tabular}[c]{@{}c@{}}Traversal \\ Weighted \\ Mean {[}m{]}\end{tabular}} &
  \textbf{\begin{tabular}[c]{@{}c@{}}Segment \\ Weighted \\ Mean {[}m{]}\end{tabular}} &
  \textbf{\begin{tabular}[c]{@{}c@{}}Traversal \\ Weighted \\ Median {[}m{]}\end{tabular}} &
  \textbf{\begin{tabular}[c]{@{}c@{}}Segment \\ Weighted \\ Median {[}m{]}\end{tabular}} &
  \textbf{\begin{tabular}[c]{@{}c@{}}\% Segments \\ w/ spacing \textgreater \\ threshold \\ (2km)\end{tabular}} \\ \hline
\multicolumn{1}{|l|}{Winnipeg Transit} &
  Winnipeg, MB &
  263 &
  274 &
  219 &
  226 &
  0.192 \\ \hline
\multicolumn{1}{|l|}{Societe   de transport de Montreal} &
  Montreal, QC &
  281 &
  308 &
  240 &
  251 &
  0.532 \\ \hline
\multicolumn{1}{|l|}{Toronto Transit Commission} &
  Toronto, ON &
  322 &
  337 &
  271 &
  273 &
  0.437 \\ \hline
\multicolumn{1}{|l|}{Reseau de transport de la Capitale} &
  Quebec, QC &
  325 &
  304 &
  285 &
  263 &
  0.559 \\ \hline
\multicolumn{1}{|l|}{Edmonton Transit Service} &
  Edmonton, AB &
  330 &
  341 &
  272 &
  282 &
  1.116 \\ \hline
\multicolumn{1}{|l|}{Ottawa-Carleton Regional Transit   Commission (OC Transpo)} &
  Ottawa, ON &
  333 &
  351 &
  261 &
  273 &
  2.164 \\ \hline
\multicolumn{1}{|l|}{Calgary Transit} &
  Calgary, AB &
  373 &
  390 &
  309 &
  318 &
  1.541 \\ \hline
\multicolumn{1}{|l|}{MiWay} &
  Mississauga, ON &
  374 &
  377 &
  303 &
  305 &
  1.004 \\ \hline
\multicolumn{1}{|l|}{TransLink Vancouver} &
  Vancouver, BC &
  378 &
  396 &
  308 &
  322 &
  1.468 \\ \hline
\multicolumn{1}{|l|}{Brampton Transit} &
  Brampton, ON &
  426 &
  406 &
  327 &
  312 &
  0.632 \\ \hline
\multicolumn{2}{|c|}{\textbf{Summary}} &
  \textbf{333} &
  \textbf{346} &
  \textbf{275} &
  \textbf{281} &
  \textbf{0.973} \\ \hline
\end{tabular}%
    }
\end{table}


Table \ref{tab:world_cities1} reports mean spacings for 38 cities outside the US and Canada. The list of cities in Table \ref{tab:world_cities1} is very much a \emph{convenience sample}. We were limited by what providers' GTFS files were possible to obtain, suitable for comparison and published with the optional `shapes.txt' file needed to calculate spacings. Obstacles to obtaining international comparisons are discussed further in Section \ref{sec:discussion}.

\begin{table}[ht]
    \caption{Stop Spacings for cities worldwide (Threshold - 2km).}\label{tab:world_cities1}
    \resizebox{\columnwidth}{!}{%
        \begin{tabular}{|p{7cm}|l|l|c|c|c|c|c|}
\hline
\multicolumn{1}{|l|}{\textbf{Provider}} &
    \textbf{\begin{tabular}[c]{@{}l@{}}Urbanized \\ Area\end{tabular}} &
  \multicolumn{1}{l|}{\textbf{Country}} &
  \textbf{\begin{tabular}[c]{@{}c@{}}Traversal \\ Weighted \\ Mean {[}m{]}\end{tabular}} &
  \textbf{\begin{tabular}[c]{@{}c@{}}Segment \\ Weighted \\ Mean {[}m{]}\end{tabular}} &
  \textbf{\begin{tabular}[c]{@{}c@{}}Traversal \\ Weighted \\ Median {[}m{]}\end{tabular}} &
  \textbf{\begin{tabular}[c]{@{}c@{}}Segment \\ Weighted \\ Median {[}m{]}\end{tabular}} &
  \textbf{\begin{tabular}[c]{@{}c@{}}\% Segments \\ w/ spacing \textgreater   \\ threshold \\ (2km)\end{tabular}} \\ \hline
\multicolumn{1}{|l|}{Azienda   Napoletana Mobilità} &
  \multicolumn{1}{l|}{Napoli} &
  Italy &
  293 &
  315 &
  256 &
  262 &
  0.551 \\ \hline
\multicolumn{1}{|l|}{Colectivos Buenos Aires} &
  \multicolumn{1}{l|}{Buenos Aires} &
  Argentina &
  309 &
  306 &
  256 &
  244 &
  1.007 \\ \hline
\multicolumn{1}{|l|}{Empresa de Transportes e Transito de Belo   Horizonte (BHTRANS)} &
  \multicolumn{1}{l|}{Belo Horizonte} &
  Brazil &
  325 &
  319 &
  266 &
  260 &
  0.43 \\ \hline
\multicolumn{1}{|l|}{Empresa Municipal de Transportes de   Madrid (EMT Madrid)} &
  \multicolumn{1}{l|}{Madrid} &
  Spain &
  332 &
  359 &
  293 &
  307 &
  0.651 \\ \hline
\multicolumn{1}{|l|}{Santiago DTPM} &
  \multicolumn{1}{l|}{Santiago} &
  Chile &
  339 &
  362 &
  302 &
  307 &
  0.391 \\ \hline
\multicolumn{1}{|l|}{Transports Metropolitans de Barcelona   (TMB)} &
  \multicolumn{1}{l|}{Barcelona} &
  Spain &
  346 &
  348 &
  324 &
  315 &
  0.176 \\ \hline
\multicolumn{1}{|l|}{Etufor} &
  \multicolumn{1}{l|}{Fortaleza} &
  Brazil &
  349 &
  352 &
  284 &
  271 &
  0.161 \\ \hline
\multicolumn{1}{|l|}{EMT Valencia} &
  \multicolumn{1}{l|}{València} &
  Spain &
  351 &
  341 &
  307 &
  296 &
  0.95 \\ \hline
\multicolumn{1}{|l|}{Azienda Trasporti Milanesi (ATM)} &
  \multicolumn{1}{l|}{Milan} &
  Italy &
  357 &
  415 &
  312 &
  343 &
  0.286 \\ \hline
\multicolumn{1}{|l|}{Dublin Bus} &
  \multicolumn{1}{l|}{Dublin} &
  Ireland &
  360 &
  414 &
  315 &
  337 &
  0.408 \\ \hline
\multicolumn{1}{|l|}{Toei Bus} &
  \multicolumn{1}{l|}{Tokyo} &
  Japan &
  364 &
  372 &
  335 &
  338 &
  0.057 \\ \hline
\multicolumn{1}{|l|}{Azienda Tramvie e Autobus del Comune di   Roma} &
  \multicolumn{1}{l|}{Rome} &
  Italy &
  369 &
  407 &
  306 &
  321 &
  0.52 \\ \hline
\multicolumn{1}{|l|}{SBS Transit (SBST)} &
  \multicolumn{1}{l|}{Singapore} &
  Singapore &
  369 &
  373 &
  345 &
  340 &
  0.943 \\ \hline
\multicolumn{1}{|l|}{Tours Métropole Val de Loire} &
  \multicolumn{1}{l|}{Tours} &
  France &
  377 &
  415 &
  336 &
  348 &
  0.274 \\ \hline
\multicolumn{1}{|l|}{Adelaide Metro} &
  \multicolumn{1}{l|}{Adelaide} &
  Australia &
  378 &
  413 &
  335 &
  350 &
  1.045 \\ \hline
\multicolumn{1}{|l|}{Wiener Lokalbahnen (WLB), Wiener Linien} &
  \multicolumn{1}{l|}{Vienna} &
  Austria &
  378 &
  414 &
  345 &
  370 &
  0.007 \\ \hline
\multicolumn{1}{|l|}{Metro Christchurch} &
  \multicolumn{1}{l|}{Christchurch} &
  New Zealand &
  379 &
  401 &
  329 &
  338 &
  0.565 \\ \hline
\multicolumn{1}{|l|}{City Transport Secretary of Rio de   Janeiro} &
  \multicolumn{1}{l|}{Rio de Janeiro} &
  Brazil &
  379 &
  427 &
  312 &
  333 &
  1.022 \\ \hline
\multicolumn{1}{|l|}{dBus} &
  \multicolumn{1}{l|}{San Sebastián} &
  Spain &
  392 &
  399 &
  345 &
  344 &
  0.061 \\ \hline
\multicolumn{1}{|l|}{Turku region public transport Föli} &
  \multicolumn{1}{l|}{Turku} &
  Finland &
  401 &
  521 &
  350 &
  430 &
  0.427 \\ \hline
\multicolumn{1}{|l|}{STIB / MIVB} &
  \multicolumn{1}{l|}{Bruxelles} &
  Belgium &
  402 &
  415 &
  379 &
  389 &
  0.188 \\ \hline
\multicolumn{1}{|l|}{Tisséo} &
  \multicolumn{1}{l|}{Toulouse} &
  France &
  407 &
  464 &
  354 &
  381 &
  0.646 \\ \hline
\multicolumn{1}{|l|}{Carris Metropolitana} &
  \multicolumn{1}{l|}{Lisboa} &
  Portugal &
  410 &
  453 &
  334 &
  363 &
  1.155 \\ \hline
\multicolumn{1}{|l|}{Berliner Verkehrsbetriebe} &
  \multicolumn{1}{l|}{Berlin} &
  Germany &
  412 &
  433 &
  376 &
  389 &
  0.395 \\ \hline
\multicolumn{1}{|l|}{Kochi buses (KSRTC)} &
  \multicolumn{1}{l|}{Kochi} &
  India &
  455 &
  502 &
  405 &
  424 &
  0.123 \\ \hline
\multicolumn{1}{|l|}{Transport Canberra} &
  \multicolumn{1}{l|}{Canberra} &
  Australia &
  465 &
  438 &
  370 &
  346 &
  3.21 \\ \hline
\multicolumn{1}{|l|}{Aachener Straßenbahn und   Energieversorgungs} &
  \multicolumn{1}{l|}{Aachen} &
  Germany &
  466 &
  512 &
  415 &
  433 &
  1.291 \\ \hline
\multicolumn{1}{|l|}{Budapesti Közlekedési Központ (BKK)} &
  \multicolumn{1}{l|}{Budapest} &
  Hungary &
  476 &
  498 &
  422 &
  433 &
  1.079 \\ \hline
\multicolumn{1}{|l|}{Gemeente Vervoerbedrijf (GVB)} &
  \multicolumn{1}{l|}{Amsterdam} &
  Netherlands &
  489 &
  525 &
  416 &
  445 &
  1.425 \\ \hline
\multicolumn{1}{|l|}{KVB Kölner Verkehrs - Betriebe AG} &
  \multicolumn{1}{l|}{Cologne} &
  Germany &
  508 &
  540 &
  453 &
  470 &
  0.977 \\ \hline
\multicolumn{1}{|l|}{Warszawski Transport Publiczny (ZTM   Warszawa)} &
  \multicolumn{1}{l|}{Warsaw} &
  Poland &
  516 &
  580 &
  458 &
  502 &
  0.609 \\ \hline
\multicolumn{1}{|l|}{MPK Wroclaw} &
  \multicolumn{1}{l|}{Wroclaw} &
  Poland &
  532 &
  625 &
  477 &
  528 &
  1.008 \\ \hline
\multicolumn{1}{|l|}{Klaipėda Transport} &
  \multicolumn{1}{l|}{Klaipėda} &
  Lithuania &
  556 &
  701 &
  505 &
  589 &
  1.177 \\ \hline
\multicolumn{1}{|l|}{Kauno viešasis transportas (KVT)} &
  \multicolumn{1}{l|}{Kaunas} &
  Lithuania &
  594 &
  644 &
  557 &
  588 &
  0.828 \\ \hline
\multicolumn{1}{|l|}{Rīgas Satiksme} &
  \multicolumn{1}{l|}{Rīga} &
  Latvia &
  596 &
  620 &
  520 &
  539 &
  1.024 \\ \hline
\multicolumn{1}{|l|}{Helsingin seudun liikenne (HSL)} &
  \multicolumn{1}{l|}{Helsinki} &
  Finland &
  611 &
  484 &
  555 &
  404 &
  1.712 \\ \hline
\multicolumn{1}{|l|}{Zarząd Transportu Publicznego w Krakowie   (ZTP Kraków)} &
  \multicolumn{1}{l|}{Kraków} &
  Poland &
  616 &
  674 &
  550 &
  604 &
  1.528 \\ \hline
\multicolumn{1}{|l|}{Vilnius Transport} &
  \multicolumn{1}{l|}{Vilnius} &
  Lithuania &
  629 &
  672 &
  545 &
  596 &
  1.713 \\ \hline
\multicolumn{3}{|c|}{\textbf{Summary}} &
  \textbf{359} &
  \textbf{403} &
  \textbf{308} &
  \textbf{337} &
  \textbf{0.774} \\ \hline
\end{tabular}%
    }
\end{table}

\clearpage
\section{Discussion}\label{sec:discussion}


\subsection{Comparisons}

As pointed out in the introduction, communication around stop consolidation has often compared average spacings among cities (even if there have not been clear methodologies or definitions for how to do so). Thus, the most significant finding of our research is probably that US bus real spacings are generally wider than the 160-230 m range (7-10 stops per mile) offered by \citet{reillyTransitServiceDesign1997} or the 200 m offered by \citet{NAP10110}. SEPTA (Philadelphia) and CTA (Chicago) are the only large US systems with traversal-weighted mean spacings shorter than 230 m. This is so even though we have excluded all spacings longer than 2 km. The traversal-weighted mean spacing for United States and Canada (truncated from below at 2 km) are even wider: about 350 m (about 5 stops per mile). The national medians are lower: between 5 and 6 stops per mile. 

A cursory look at city-level data suggests that cities largely built before the automobile area have narrower spacings than those which have developed since. In the US, Philadelphia, Chicago, Boston, Pittsburgh and San Francisco have much shorter-than-average spacings while Las Vegas, Austin and Santa Clara County have much wider spacings. Similarly, in Canada, Mississauga and Brampton have wider spacings than their immediate neighbor Toronto. The relationship between stop spacing, urban form and development history is a question worth investigating in further research.

\begin{figure}[ht]
    \centering
    \includegraphics[width =.7\textwidth]{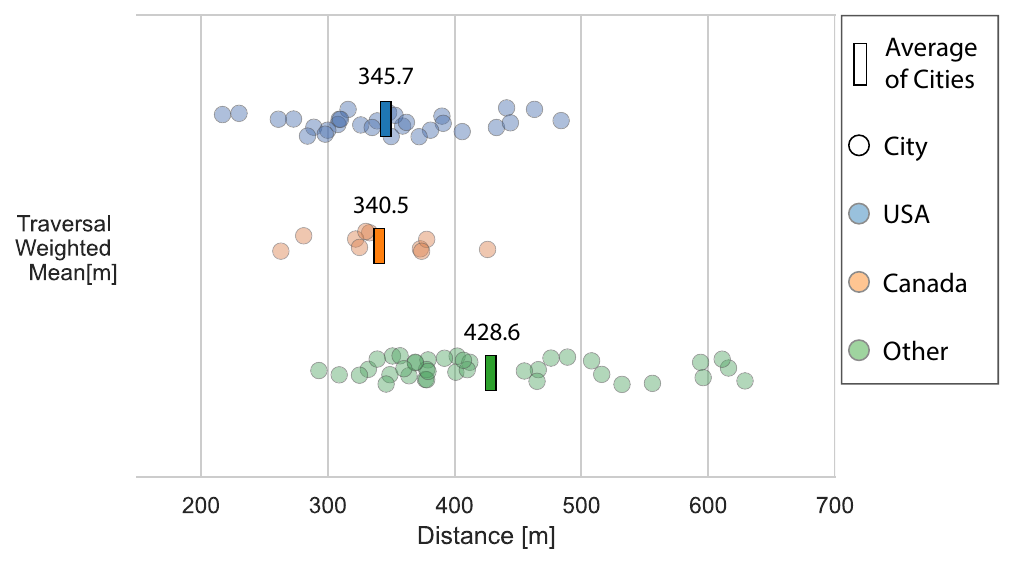}
    \caption{Comparison of Stop Spacings between US, Canada, and other countries }
    \label{fig:strip_plot}
\end{figure}

While US cities generally have wider spacings than thought, in another way the conventional wisdom (as reflected by the three quotes given in the introduction) seems to be right. In our sample of 38 international cities, spacings abroad are generally wider than in the US (and Canada). The mean spacing in the international sample is wider than all ten Canadian systems, and all but five of the thirty largest US systems. Figure \ref{fig:strip_plot} displays a ``strip plot'' for the 78 cities displayed in Tables \ref{tab:US_cities}, \ref{tab:CA_cities} and \ref{tab:world_cities1}. Each city's spacing appears as a dot. Among cities in the international sample, Eastern Europe (Poland and the Baltic countries) have the widest spacings---above 500 meters (just fewer than three stops per mile).

Still, while the international data are suggestive, we would caution against drawing confident conclusions from the data depicted in Table \ref{tab:world_cities1}. Our databases contain hundreds of cities in the US and Canada, but data from other countries proved much harder to obtain for two major reasons. First (and most importantly), providers outside the US and Canada\footnote{Inquiring with Google employees, we learned that navigation apps such as Google Maps generally use a combination of real-time GPS data and manual operations to draw bus route shapes.} rarely include the optional shapes.txt file needed to determine driving distances between stops. There is little relationship between an agency's size and whether it publishes a shapes.txt: e.g., Paris' Ile-de-France Mobilité does not but Toulouse's Tisséo does. The second obstacle was conceptual: many countries have no equivalents of a city bus system comparable to US and Canadian systems. In some places this is because one agency serves a large region: e.g., in Norway, Ruter AS serves Oslo and Akershus counties, including vast rural areas. In the United Kingdom, by contrast, cities are served by numerous private\footnote{Having many private providers is not fatal to spacing comparisons. For example, Buenos Aires organizes its privately-provided routes into a unified GTFS file at \url{https://data.buenosaires.gob.ar/dataset/colectivos-gtfs}} providers which operate across cities or even nationally. In either case, one could ostensibly filter all segments from within some political boundary, but we have not done so.

\subsection{Advantages of isolating segments}

\begin{figure}[ht]
    \centering
    \begin{subfigure}[b]{0.4\textwidth}
        \centering
        \includegraphics[width=\textwidth]{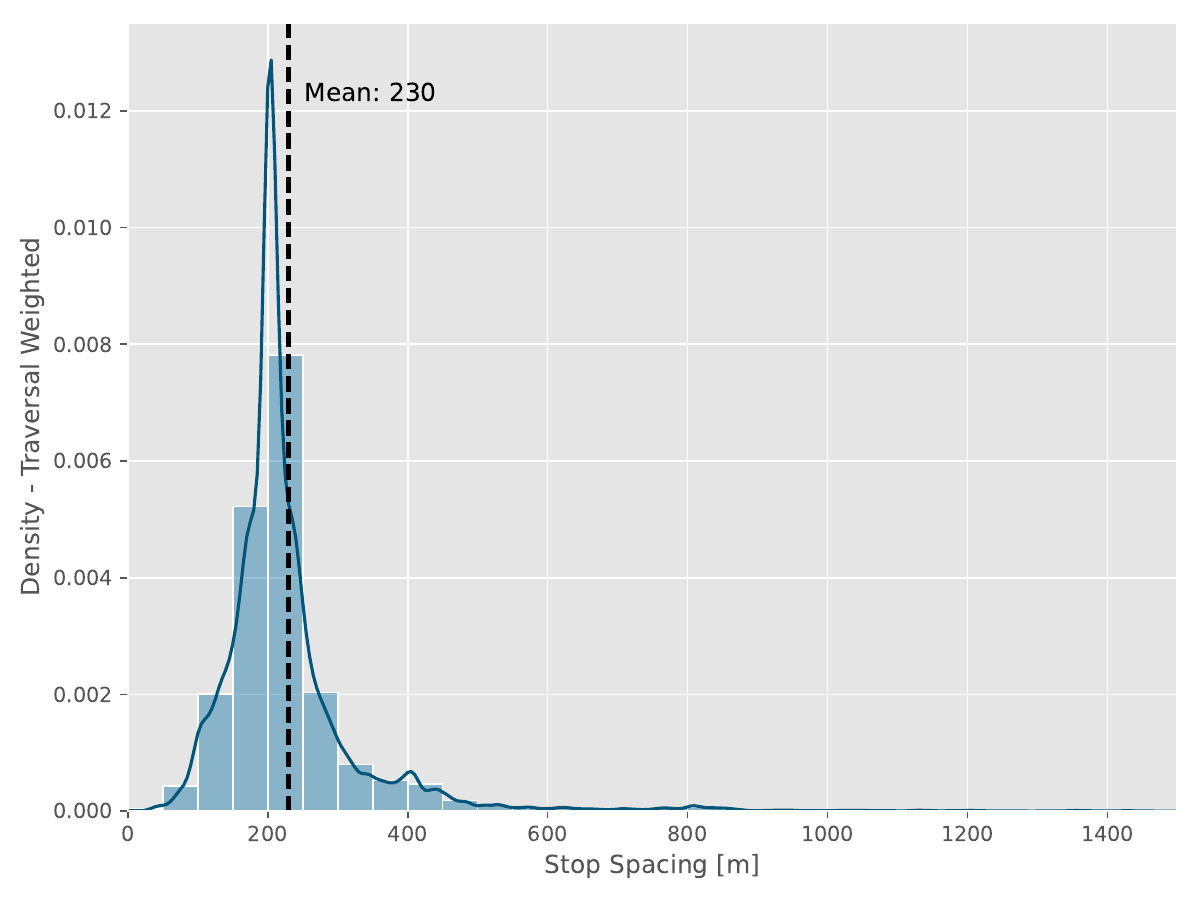}
        \caption{CTA, Chicago}
        \label{fig:chicago1}
    \end{subfigure}
    \begin{subfigure}[b]{0.4\textwidth}
        \centering
        \includegraphics[width=\textwidth]{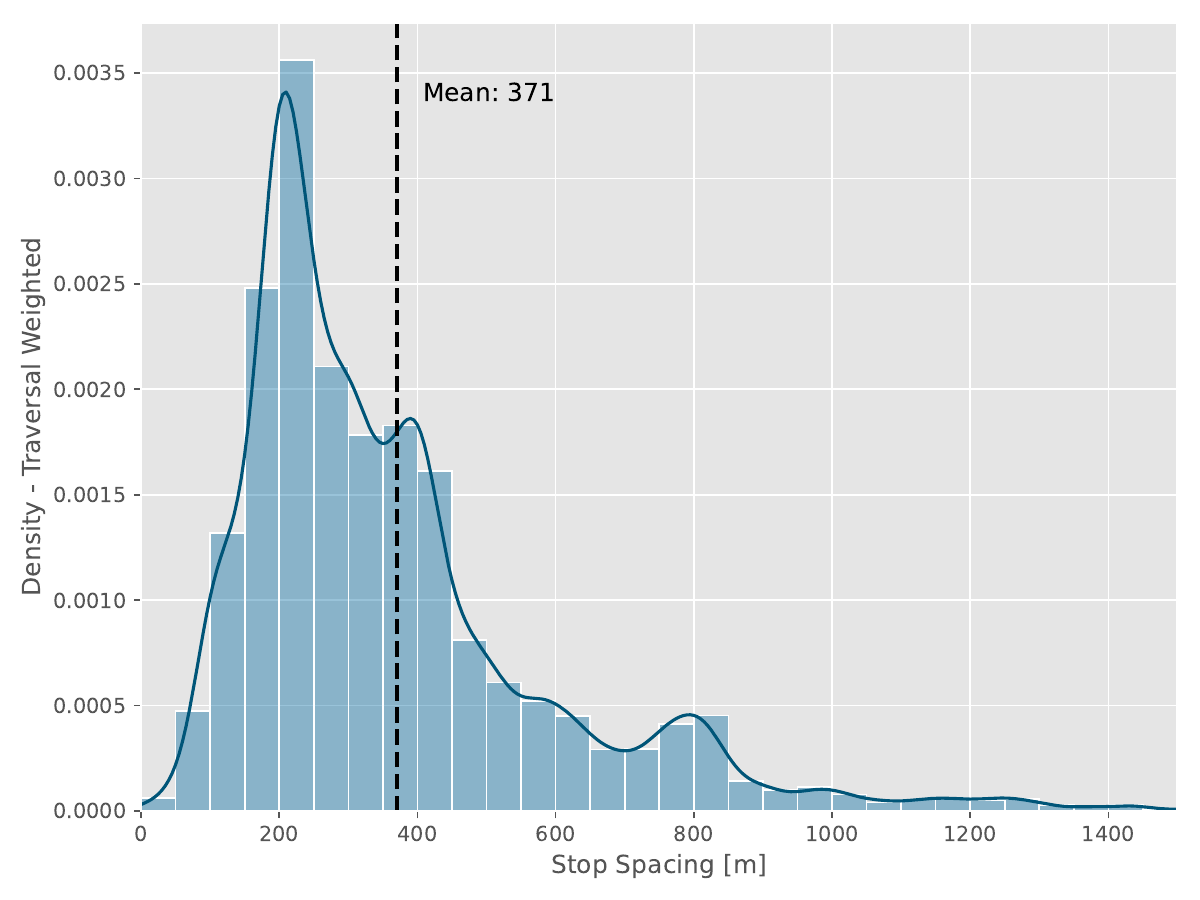}
        \caption{RideKC, Kansas City}
        \label{fig:kansas1}
    \end{subfigure}
    \begin{subfigure}[b]{0.4\textwidth}
        \centering
        \includegraphics[width=\textwidth]{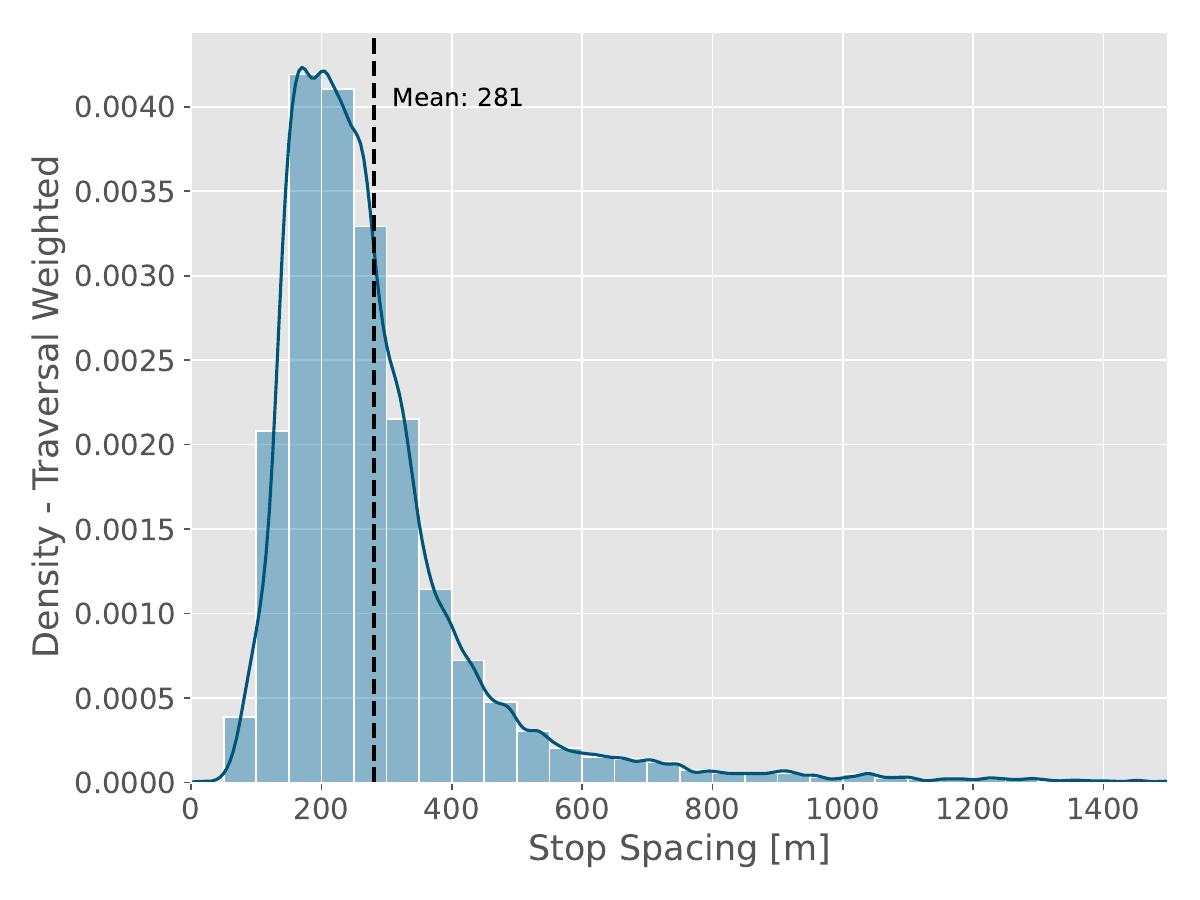}
        \caption{Société de transport de Montréal, Montreal}
        \label{fig:montreal1}
    \end{subfigure}
    \begin{subfigure}[b]{0.4\textwidth}
        \centering
        \includegraphics[width=\textwidth]{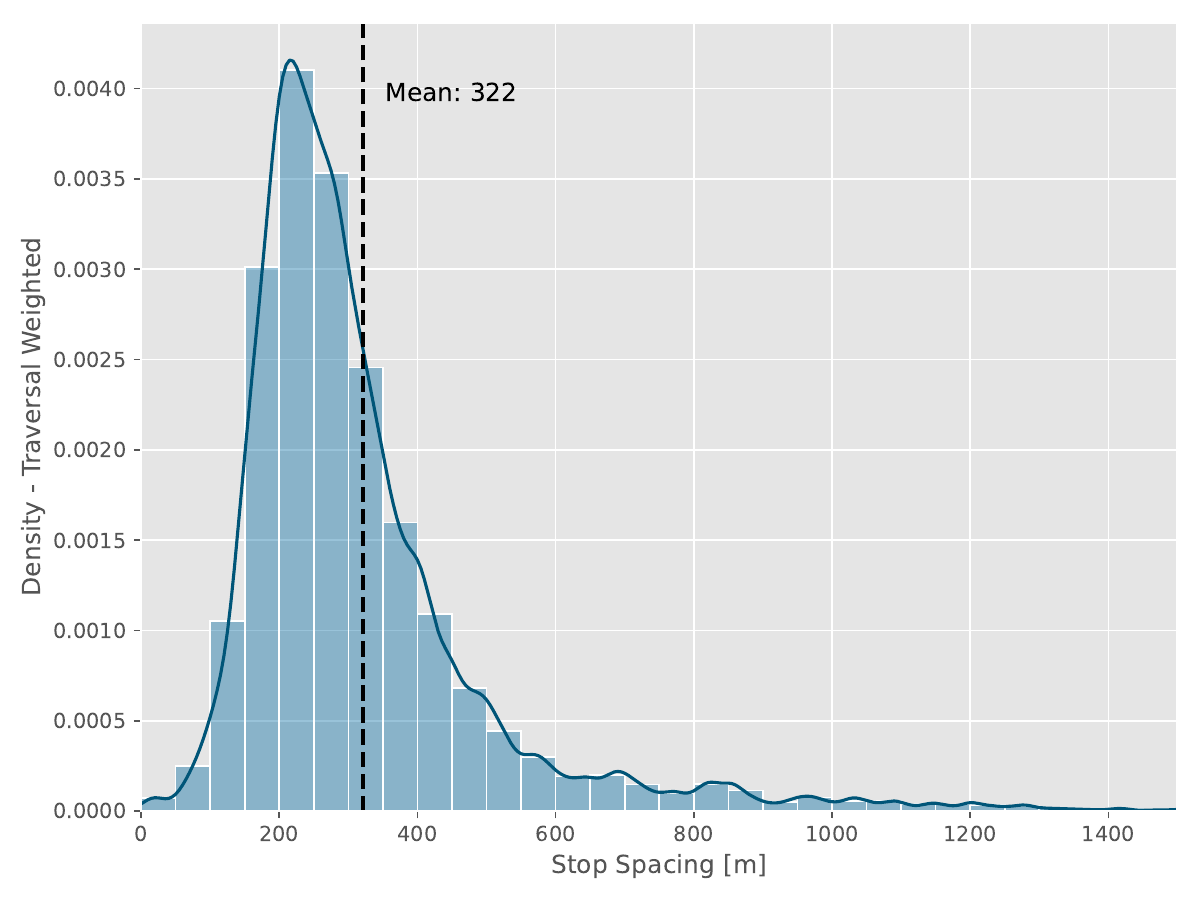}
        \caption{Toronto Transit Commission, Toronto}
        \label{fig:toronto1}
    \end{subfigure}
    \caption{Traversal-weighted distributions of stop spacings}
    \label{fig:dists}
\end{figure}

In Section \ref{sec:methodology}, we described a methodology and introduced a python package for splitting bus networks into segments. However, it is possible to obtain some means \emph{without} isolating segments. The average spacing of each trip in the schedule can be calculated by measuring the distance of each trip's shape and then dividing that distance by the number of stops on the trip. By averaging the average spacing of all trips, one obtains a traversal-weighted spacing. By counting the average spacing of a trip on each route once, one obtains a route-weighted spacing. What advantages, then, are there to isolating segments individually? Isolating segments permits: 
\begin{enumerate}[label=(\roman*)]
    \item statistics which by their nature depend on counting individual spacings, such as medians, quartiles, standard deviations and the segment-weighted mean;
    \item particular spacings to be filtered (e.g., to discard segments which are longer than a threshold or which fall outside some political boundary); 
    \item the creation of histograms of the complete distribution of spacing. 
\end{enumerate}
Figure \ref{fig:dists} shows spacing distributions for traversal-weighted histograms and kernel densities of stop spacings for Chicago, Kansas City, Montreal, and Toronto's bus systems. In every city we have examined, the distribution has a similar shape: unimodal with a right skew. 

\subsection{Alternative metrics}\label{sec:extension}

Before concluding, we now consider two alternative ways to calculate stop spacings. These are intended as proofs-of-concepts showing what can be accomplished by combining GTFS data with other data sources.

\begin{figure}[ht]
    \centering
    \begin{subfigure}[b]{0.40\textwidth}
        \centering
        \includegraphics[width=\textwidth]{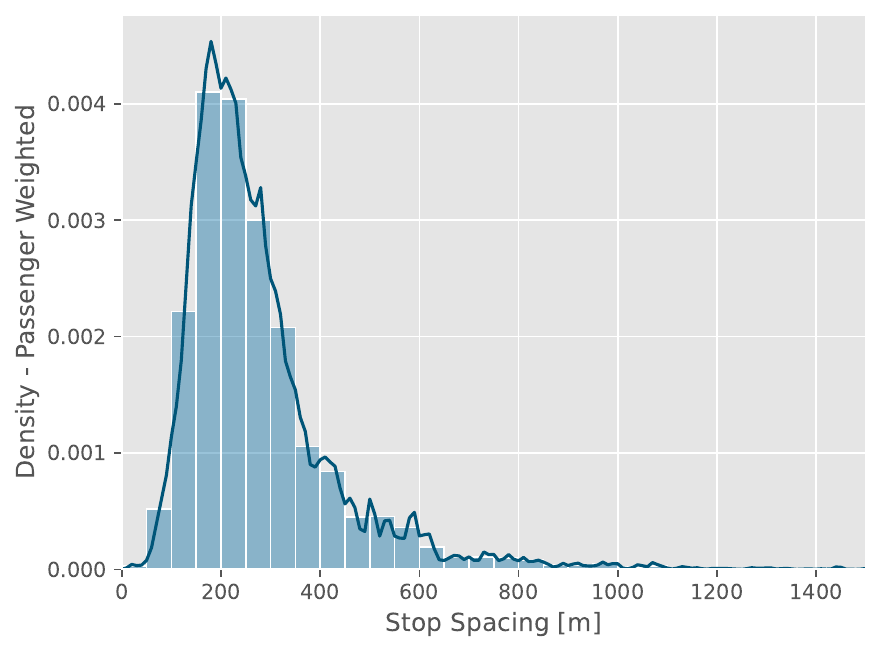}
        \caption{MBTA, Boston}
        \label{fig:boston1}
    \end{subfigure}
    \begin{subfigure}[b]{0.40\textwidth}
        \centering
        \includegraphics[width=\textwidth]{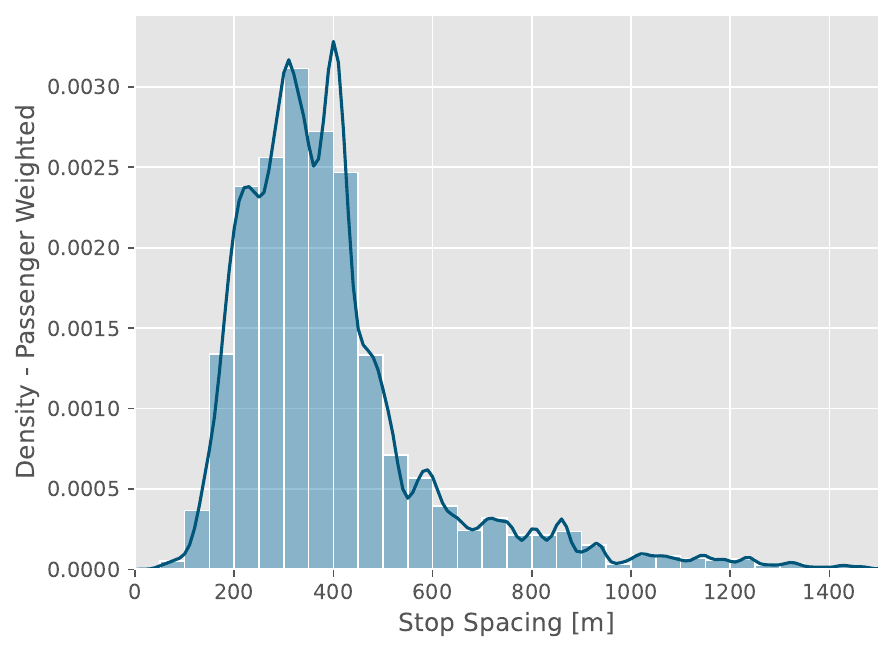}
        \caption{RTD, Denver}
        \label{fig:denver1}
    \end{subfigure}
    \caption{Distributions of passenger-weighted stop spacings}
    \label{fig:load_dists}
\end{figure}

The first is a new weighting scheme: \emph{passenger-weighting}. By this scheme, segments are weighted by the average number of passengers who traverse the segment. Just as the traversal-weighted mean is the average spacing ``experienced'' by a bus, the passenger-weighted mean is the average spacing experienced by a passenger. The MBTA (in Boston) publishes the bus loads for a large sample of trips on their Open Data Portal \citep{MassachusettsBayTransportationAuthority2019}. Upon request, we obtained the same data from RTD in Denver. Figure \ref{fig:load_dists} shows each agency's passenger-weighted distribution. The  MBTA's passenger-weighted mean is 304 m; RTD's is 401 m. However, neither agency samples all scheduled trips, so these means are \emph{not comparable} to the other means reported. It would be possible, however, to obtain comparable means by interpolating loads for missing trips.

\begin{figure}[ht]
    \centering
    \begin{subfigure}[b]{0.40\textwidth}
        \centering
        \includegraphics[width=\textwidth]{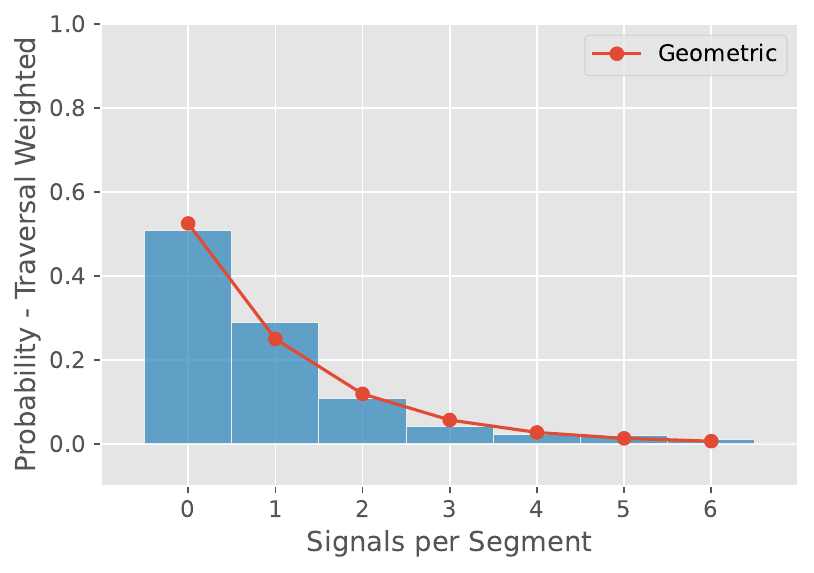}
        \caption{Austin, TX}
        \label{fig:austin}
    \end{subfigure}
    \begin{subfigure}[b]{0.40\textwidth}
        \centering
        \includegraphics[width=\textwidth]{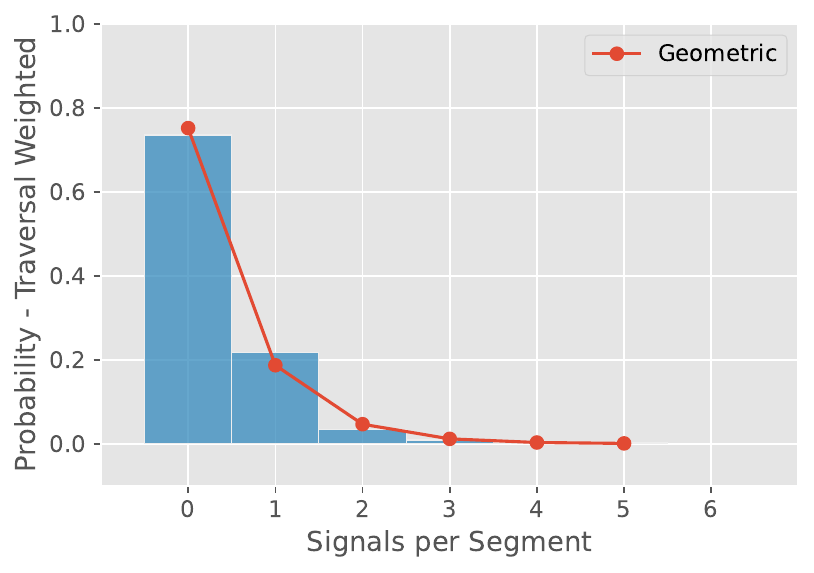}
        \caption{Bloomington, IN}
        \label{fig:Bloomington}
    \end{subfigure}
    \begin{subfigure}[b]{0.40\textwidth}
        \centering
        \includegraphics[width=\textwidth]{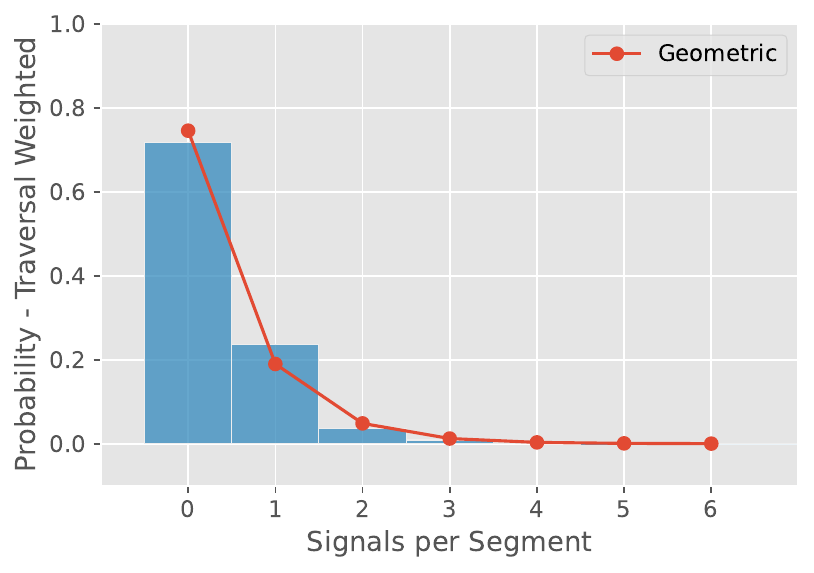}
        \caption{Chicago, IL}
        \label{fig:chicago}
    \end{subfigure}
    \begin{subfigure}[b]{0.40\textwidth}
        \centering
        \includegraphics[width=\textwidth]{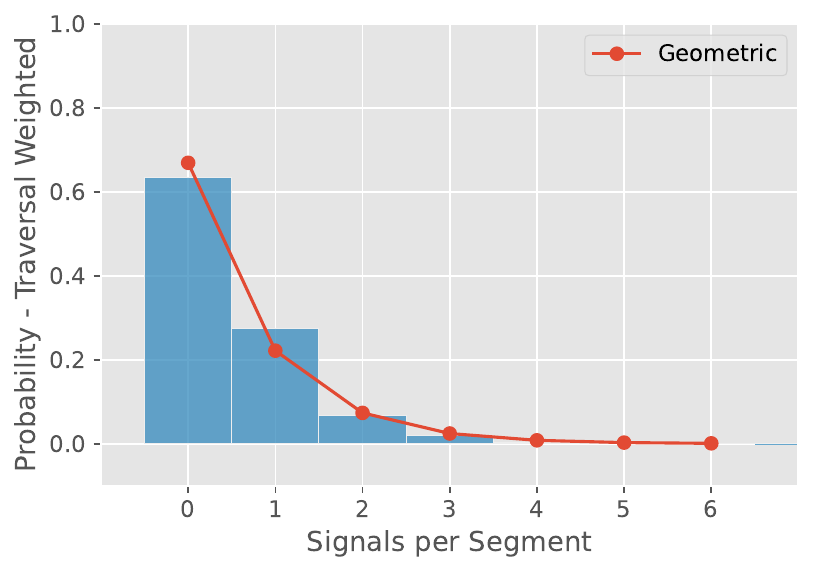}
        \caption{Pittsburgh, PA}
        \label{fig:pitsburgh}
    \end{subfigure}
    \caption{Traversal-weighted distributions of signals-per-segment}
    \label{fig:signals}
\end{figure}

The next proof-of-concept is an alternative distance metric: traffic signals traversed. The number of traffic signals, while usually not under the control of a bus agency, also communicates information about how much time it takes the bus to travel between stops. From various government sources, we have obtained data that includes the coordinates of traffic signals in twelve cities. To calculate signals-traversed, we find the number of signals falling within a 5.5 m buffer around each segment. Table \ref{tab:average-signal} shows segment-, route- and traversal-weighted mean values of signals-per-segment for each of the twelve cities. Figure \ref{fig:signals} shows the full traversal-weighted distributions of signals-per-segment. Most segments have no signal, but the number can be as large as six in Austin. Using Maximum Likelihood Estimation, we fit a geometric distribution to each data, with the predicted probability masses shown in red in Figure \ref{fig:signals}. The geometric distribution fits surprisingly well.

\begin{table}[ht]
    \centering
    \caption{Average signals-per-segment for 12 US cities}\label{tab:average-signal}
    \resizebox{.8\columnwidth}{!}{%
        \begin{tabular}{|l|l|c|c|c|c|}
  \hline
  \textbf{Provider}                                                                                            & \textbf{\begin{tabular}[c]{@{}l@{}}Urbanized \\ Area\end{tabular}} & \multicolumn{1}{l|}{\textbf{State}} & \multicolumn{1}{l|}{\textbf{\begin{tabular}[c]{@{}l@{}}Segment-\\ Weighted \\ Mean\end{tabular}}} & \multicolumn{1}{c|}{\textbf{\begin{tabular}[c]{@{}l@{}}Route-\\ Weighted \\ Mean\end{tabular}}} & \multicolumn{1}{c|}{\textbf{\begin{tabular}[c]{@{}l@{}}Traversal-\\ Weighted \\ Mean\end{tabular}}} \\ \hline
  Capital Metro                                                                                                & Austin                                                             & TX                                  & 0.83                                                                                              & 0.91                                                                                            & 0.91                                                                                                \\ \hline
  Bloomington Transit                                                                                          & Bloomington                                                        & IN                                  & 0.32                                                                                              & 0.33                                                                                            & 0.33                                                                                                \\ \hline
  Chicago Transit Authority (CTA)                                                                              & Chicago                                                            & IL                                  & 0.31                                                                                              & 0.32                                                                                            & 0.34                                                                                                \\ \hline
  \begin{tabular}[c]{@{}l@{}}Jacksonville Transportation \\ Authority (JTA)\end{tabular}                       & Jacksonville                                                       & FL                                  & 0.39                                                                                              & 0.40                                                                                            & 0.45                                                                                                \\ \hline
  \begin{tabular}[c]{@{}l@{}}Transit Authority of River \\ City (TARC)\end{tabular}                            & Louisville                                                         & KY                                  & 0.47                                                                                              & 0.53                                                                                            & 0.68                                                                                                \\ \hline
  Miami-Dade Transit                                                                                           & Miami                                                              & FL                                  & 0.37                                                                                              & 0.37                                                                                            & 0.41                                                                                                \\ \hline
  \begin{tabular}[c]{@{}l@{}}Central Florida Regional \\ Transit Authority (LYNX)\end{tabular}                 & Orlando                                                            & FL                                  & 0.23                                                                                              & 0.23                                                                                            & 0.22                                                                                                \\ \hline
  \begin{tabular}[c]{@{}l@{}}Southeastern Pennsylvania \\ Transportation Authority (SEPTA)\end{tabular}        & Philadelphia                                                       & PA                                  & 0.63                                                                                              & 0.61                                                                                            & 0.61                                                                                                \\ \hline
  Pittsburgh Regional Transit                                                                                  & Pittsburgh                                                         & PA                                  & 0.33                                                                                              & 0.41                                                                                            & 0.50                                                                                                \\ \hline
  \begin{tabular}[c]{@{}l@{}}San Francisco Municipal \\ Transportation Agency (SFMTA)\end{tabular}             & San Francisco                                                      & CA                                  & 1.06                                                                                              & 1.14                                                                                            & 1.26                                                                                                \\ \hline
  \begin{tabular}[c]{@{}l@{}}Hillsborough Area Regional \\ Transit (HART)\end{tabular}                         & Tampa                                                              & FL                                  & 0.28                                                                                              & 0.28                                                                                            & 0.33                                                                                                \\ \hline
  King County Metro
  & Seattle                                                         & WA                                  & 0.38                                                                                              & 0.56                                                                                            & 0.66                                                                                                \\ \hline
\end{tabular}%
    }
\end{table}

\clearpage
\section{Conclusion}\label{sec:conclusion}

US transit agencies are currently reconsidering their stop policies, and discussions around these policies sometimes reference ``typical'' or ``average'' values for comparison. One task of the paper has been to provide conceptual clarity to discussions of stop spacing statistics by introducing terminologies and weighting schemes. Another task has been to describe a methodology (implemented via the python package \emph{gtfs-segments}) for isolating individual segments from GTFS data. The concepts and tools together were used to provide spacing statistics for the United States and Canada nationally, as well as thirty large US providers, ten Canadian ones and thirty-eight in other countries. Evidence shows that US (and Canadian) spacings are not quite as narrow as some quoted statistics suggest, but do seem to be somewhat narrower than those in other countries---although the sample of international providers is not large enough to make confident comparisons. Finally, the paper presented, as proofs-of-concept, two spacing statistics that combine GTFS with outside data sources.

In any discussion of stop spacings, a note of caution is in order: the very idea of comparing spacing statistics should be taken with a grain of salt. For example, Chicago's mean spacings would be wider if the Chicago Transit Authority (which operates only in the dense City of Chicago) and Pace (which operates in Chicago's suburbs) were a single agency. Still, both agencies do have average spacings shorter than the US mean, so aggregates do reflect ``composition effects'' but also provide (at least suggestive) evidence about genuine policy differences. Even though the paper is premised on the idea that statistics are useful, one very general point which the authors hope the paper has made clear is how \emph{synthetic} any particular statistic is---dependent on choices, weightings, exclusions, etc. Hence, it is of foremost importance to be precise about how exactly aggregates were aggregated.

To conclude, we propose ideas for further work. One promising avenue for future research is to incorporate the GTFS Realtime specification, which supplies live bus location information online. One use of this data would be to compute metrics of \emph{realized} bus stop spacings. While GTFS Schedule data records the locations of stops, typically buses only actually visit a stop if a passenger wants to alight or if there is someone waiting at the stop. By collecting real-time bus locations, it would be possible to record when buses actually stop and thereby measure the actual distances traveled from one stopping operation to another. This is important to understanding stop removal, because on low-demand routes where buses only actually visit a small fraction of stops (as in smaller communities), spacings may greatly underestimate how far the bus can travel freely. Similarly, GTFS Realtime data could be used to measure (weighted in various ways) \emph{travel time between stops} as a distance metric.

The ideas and tools presented in this paper can also be used for more general research. One way to proceed is to use stop-by-stop passenger data, such as that the passenger count data provided by MBTA and RTD in Section \ref{sec:extension}, to identify relationships between changes in ridership and changes in bus stop spacing. When a stop is removed, do all passengers use nearby stops, or do any trips vanish? Do the quicker bus speeds (and, potentially, shorter headways) afforded by consolidation boost demand on consolidated routes? Because GTFS and ridership data can be acquired over time, it should be possible to use time-series techniques such as regression discontinuity and difference-in-differences to answer such questions.

Finally, the fine-grained data provided by our methodology is well-suited to geospatial analysis of stop changes. One might identify local correlates of stop removal, by linking schedule information to demographic and geographic data. Which types of locations and neighborhoods tend to have stops removed? Or, apart from removal, one might simply identify correlates of current spacings. For example, many US agencies seem to put a stop on every or every other block in urban settings, and in this case stop spacing should be highly correlated with block length. Time will tell whether these or other questions are of interest to the research community, but we offer them as examples of how data on isolated stop spacings may be used to address broader topics.



\clearpage
\bibliographystyle{elsarticle-harv}
\bibliography{bib}

\begin{thebibliography}{38}
\expandafter\ifx\csname natexlab\endcsname\relax\def\natexlab#1{#1}\fi
\providecommand{\url}[1]{\texttt{#1}}
\providecommand{\href}[2]{#2}
\providecommand{\path}[1]{#1}
\providecommand{\DOIprefix}{doi:}
\providecommand{\ArXivprefix}{arXiv:}
\providecommand{\URLprefix}{URL: }
\providecommand{\Pubmedprefix}{pmid:}
\providecommand{\doi}[1]{\href{http://dx.doi.org/#1}{\path{#1}}}
\providecommand{\Pubmed}[1]{\href{pmid:#1}{\path{#1}}}
\providecommand{\bibinfo}[2]{#2}
\ifx\xfnm\relax \def\xfnm[#1]{\unskip,\space#1}\fi
\bibitem[{Berez(2015)}]{berezBusStopsHere2015}
\bibinfo{author}{Berez, D.}, \bibinfo{year}{2015}.
\newblock \bibinfo{title}{The {{Bus Stops Here}}: {{Best Practices}} in {{Bus Stop Consolidation}} and {{Optimization}}}.
\newblock \bibinfo{publisher}{{eScholarship, University of California Los Angeles}}, \bibinfo{address}{{Los Angeles}}.
\newblock \URLprefix \url{https://escholarship.org/content/qt4dd8h1vs/qt4dd8h1vs.pdf}.
\bibitem[{Blazina(2020)}]{blazinaPortAuthorityInitial2020}
\bibinfo{author}{Blazina, E.}, \bibinfo{year}{2020}.
\newblock \bibinfo{title}{Port {{Authority}}'s initial bus stop eliminations showing on-time improvements}.
\newblock \bibinfo{journal}{Pittsburgh Post-Gazette} \URLprefix \url{https://www.post-gazette.com/news/transportation/2020/02/23/Port-Authority-bus-stops-on-time-performance-improvements-efficiency-transit/stories/202002230032}.
\bibitem[{Devunuri(2022)}]{gtfs_segments}
\bibinfo{author}{Devunuri, S.}, \bibinfo{year}{2022}.
\newblock \bibinfo{title}{gtfs-segments - a fast and efficient library to generate bus stop spacings}.
\newblock \bibinfo{journal}{Github} .
\bibitem[{Devunuri et~al.(2022)Devunuri, Qiam and Lehe}]{Devunuri2022}
\bibinfo{author}{Devunuri, S.}, \bibinfo{author}{Qiam, S.}, \bibinfo{author}{Lehe, L.}, \bibinfo{year}{2022}.
\newblock \bibinfo{title}{{Bus Stop Spacings for Transit Providers in the US - Harvard Dataverse}}.
\newblock \bibinfo{howpublished}{\url{https://dataverse.harvard.edu/dataset.xhtml?persistentId=doi:10.7910/DVN/SFBIVU}}.
\bibitem[{Devunuri et~al.(2023)Devunuri, Qiam and Lehe}]{Devunuri2023}
\bibinfo{author}{Devunuri, S.}, \bibinfo{author}{Qiam, S.}, \bibinfo{author}{Lehe, L.}, \bibinfo{year}{2023}.
\newblock \bibinfo{title}{{Bus Stop Spacings for Transit Providers in Canada}}.
\newblock \URLprefix \url{https://doi.org/10.7910/DVN/QFTAPM}, \DOIprefix\doi{10.7910/DVN/QFTAPM}.
\bibitem[{{El-Geneidy} et~al.(2006){El-Geneidy}, Strathman, Kimpel and Crout}]{el-geneidyEffectsBusStop2006}
\bibinfo{author}{{El-Geneidy}, A.M.}, \bibinfo{author}{Strathman, J.G.}, \bibinfo{author}{Kimpel, T.J.}, \bibinfo{author}{Crout, D.T.}, \bibinfo{year}{2006}.
\newblock \bibinfo{title}{Effects of bus stop consolidation on passenger activity and transit operations}.
\newblock \bibinfo{journal}{Transportation Research Record: Journal of the Transportation} , \bibinfo{pages}{32--41}\DOIprefix\doi{10.3141/1971-06}.
\bibitem[{Erhardt et~al.(2022)Erhardt, Hoque, Goyal, Berrebi, Brakewood and Watkins}]{erhardt2022has}
\bibinfo{author}{Erhardt, G.D.}, \bibinfo{author}{Hoque, J.M.}, \bibinfo{author}{Goyal, V.}, \bibinfo{author}{Berrebi, S.}, \bibinfo{author}{Brakewood, C.}, \bibinfo{author}{Watkins, K.E.}, \bibinfo{year}{2022}.
\newblock \bibinfo{title}{Why has public transit ridership declined in the united states?}
\newblock \bibinfo{journal}{Transportation Research Part A: Policy and Practice} \bibinfo{volume}{161}, \bibinfo{pages}{68--87}.
\bibitem[{Flint et~al.(2014)Flint, {Ben-Amos}, Ellis and Krykewycz}]{Flint2014}
\bibinfo{author}{Flint, T.}, \bibinfo{author}{{Ben-Amos}, A.}, \bibinfo{author}{Ellis, P.}, \bibinfo{author}{Krykewycz, G.R.}, \bibinfo{year}{2014}.
\newblock \bibinfo{title}{Piloting {{Low-Cost Transit Service Enhancements}} through {{Agency Collaboration}}}.
\newblock \bibinfo{journal}{Transportation Research Record} \bibinfo{volume}{2416}, \bibinfo{pages}{10--18}.
\newblock \DOIprefix\doi{10.3141/2416-02}.
\bibitem[{Furth and Rahbee(2000)}]{furth2000}
\bibinfo{author}{Furth, P.G.}, \bibinfo{author}{Rahbee, A.B.}, \bibinfo{year}{2000}.
\newblock \bibinfo{title}{Dynamic {{Programming}} and {{Geographic Modeling}}}.
\newblock \bibinfo{journal}{Transportation Research Record: Journal of the Transportation} \bibinfo{volume}{00--0870}, \bibinfo{pages}{15--22}.
\bibitem[{Garnham(2020)}]{garnhamFightHugeDrop2020}
\bibinfo{author}{Garnham, J.P.}, \bibinfo{year}{2020}.
\newblock \bibinfo{title}{To fight huge drop in bus riders , {{North Texas}} transit agency faces hard choices about who gets service}.
\newblock \bibinfo{journal}{Texas Tribune} \URLprefix \url{https://www.texastribune.org/2020/02/25/dallas-bus-ridership-plummeting-so-dart-wants-redraw-bus-routes-2020/}.
\bibitem[{Gordon(2010)}]{gordonMuniMayReduce2010}
\bibinfo{author}{Gordon, R.}, \bibinfo{year}{2010}.
\newblock \bibinfo{title}{Muni may reduce stops to increase speed, save cash}.
\newblock \URLprefix \url{https://www.sfgate.com/bayarea/article/Muni-may-reduce-stops-to-increase-speed-save-cash-3168017.php}.
\bibitem[{Graehler et~al.(2019)Graehler, Mucci and Erhardt}]{graehler2019understanding}
\bibinfo{author}{Graehler, M.}, \bibinfo{author}{Mucci, R.A.}, \bibinfo{author}{Erhardt, G.D.}, \bibinfo{year}{2019}.
\newblock \bibinfo{title}{Understanding the recent transit ridership decline in major us cities: Service cuts or emerging modes}, in: \bibinfo{booktitle}{Transportation Research Board 98th Annual Meeting, Washington, DC, January}.
\bibitem[{Grahn et~al.(2021)Grahn, Qian, Matthews and Hendrickson}]{grahn2021travelers}
\bibinfo{author}{Grahn, R.}, \bibinfo{author}{Qian, S.}, \bibinfo{author}{Matthews, H.S.}, \bibinfo{author}{Hendrickson, C.}, \bibinfo{year}{2021}.
\newblock \bibinfo{title}{Are travelers substituting between transportation network companies (tnc) and public buses? a case study in pittsburgh}.
\newblock \bibinfo{journal}{Transportation} \bibinfo{volume}{48}, \bibinfo{pages}{977--1005}.
\bibitem[{Jordahl et~al.(2020)Jordahl, den Bossche, Fleischmann, Wasserman, McBride, Gerard, Tratner, Perry, Badaracco, Farmer, Hjelle, Snow, Cochran, Gillies, Culbertson, Bartos, Eubank, maxalbert, Bilogur, Rey, Ren, Arribas-Bel, Wasser, Wolf, Journois, Wilson, Greenhall, Holdgraf, Filipe and Leblanc}]{kelsey_jordahl_2020_3946761}
\bibinfo{author}{Jordahl, K.}, \bibinfo{author}{den Bossche, J.V.}, \bibinfo{author}{Fleischmann, M.}, \bibinfo{author}{Wasserman, J.}, \bibinfo{author}{McBride, J.}, \bibinfo{author}{Gerard, J.}, \bibinfo{author}{Tratner, J.}, \bibinfo{author}{Perry, M.}, \bibinfo{author}{Badaracco, A.G.}, \bibinfo{author}{Farmer, C.}, \bibinfo{author}{Hjelle, G.A.}, \bibinfo{author}{Snow, A.D.}, \bibinfo{author}{Cochran, M.}, \bibinfo{author}{Gillies, S.}, \bibinfo{author}{Culbertson, L.}, \bibinfo{author}{Bartos, M.}, \bibinfo{author}{Eubank, N.}, \bibinfo{author}{maxalbert}, \bibinfo{author}{Bilogur, A.}, \bibinfo{author}{Rey, S.}, \bibinfo{author}{Ren, C.}, \bibinfo{author}{Arribas-Bel, D.}, \bibinfo{author}{Wasser, L.}, \bibinfo{author}{Wolf, L.J.}, \bibinfo{author}{Journois, M.}, \bibinfo{author}{Wilson, J.}, \bibinfo{author}{Greenhall, A.}, \bibinfo{author}{Holdgraf, C.}, \bibinfo{author}{Filipe}, \bibinfo{author}{Leblanc, F.}, \bibinfo{year}{2020}.
\newblock \bibinfo{title}{geopandas/geopandas: v0.8.1}.
\newblock \URLprefix \url{https://doi.org/10.5281/zenodo.3946761}, \DOIprefix\doi{10.5281/zenodo.3946761}.
\bibitem[{{King County Metro}(2021)}]{kingcountymetroTransitSpeedReliability2021}
\bibinfo{author}{{King County Metro}}, \bibinfo{year}{2021}.
\newblock \bibinfo{title}{Transit {{Speed}} and {{Reliability}}: {{Guidlines}} and {{Strategies}}}.
\newblock \bibinfo{type}{Technical Report}. \bibinfo{address}{{Seattle, WA}}.
\newblock \URLprefix \url{https://kingcounty.gov/{~}/media/depts/metro/about/planning/speed-reliability-toolbox.pdf}.
\bibitem[{LaFleur(2019)}]{laFleur2019}
\bibinfo{author}{LaFleur, P.}, \bibinfo{year}{2019}.
\newblock \bibinfo{title}{Cincinnati metro program removed hundreds of bus stops over the weekend}.
\newblock \bibinfo{journal}{WCPO Cincinnati} \URLprefix \url{https://www.wcpo.com/news/transportation-development/public-transit/cincinnati-metro-program-removed-hundreds-of-bus-stops-over-the-weekend}.
\bibitem[{Li and Bertini(2009)}]{liAssessingModelOptimal2009}
\bibinfo{author}{Li, H.}, \bibinfo{author}{Bertini, R.L.}, \bibinfo{year}{2009}.
\newblock \bibinfo{title}{Assessing a model for optimal bus stop spacing with high-resolution archived stop-level data}.
\newblock \bibinfo{journal}{Transportation Research Record: Journal of the Transportation} , \bibinfo{pages}{24--32}\DOIprefix\doi{10.3141/2111-04}.
\bibitem[{{Massachusetts Bay Transportation Authority}(2019)}]{MassachusettsBayTransportationAuthority2019}
\bibinfo{author}{{Massachusetts Bay Transportation Authority}}, \bibinfo{year}{2019}.
\newblock \bibinfo{title}{{MBTA Bus Ridership by Trip, Season, Route/Line, and Stop | MBTA Blue Book Open Data Portal}}.
\newblock \bibinfo{note}{\url{https://mbta-massdot.opendata.arcgis.com/datasets/MassDOT::mbta-bus-ridership-by-trip-season-route-line-and-stop/explore?filters=eyJzZWFzb24iOlsiRmFsbCAyMDE5Il19}}.
\bibitem[{McHugh(2013)}]{mchugh2013pioneering}
\bibinfo{author}{McHugh, B.}, \bibinfo{year}{2013}.
\newblock \bibinfo{title}{Pioneering open data standards: The gtfs story}.
\newblock \bibinfo{journal}{Beyond transparency: open data and the future of civic innovation} , \bibinfo{pages}{125--135}.
\bibitem[{Miatkowski and Hovenkotter(2019)}]{miatkowskiBusStopBalancing2019}
\bibinfo{author}{Miatkowski, P.}, \bibinfo{author}{Hovenkotter, K.}, \bibinfo{year}{2019}.
\newblock \bibinfo{title}{Bus {{Stop Balancing}}: {{A Campaign Guide}} for {{Agency Staff}}}.
\newblock \bibinfo{type}{Technical Report}. {TransitCenter}. \bibinfo{address}{{New York, NY}}.
\newblock \URLprefix \url{https://transitcenter.org/wp-content/uploads/2019/07/BusStopBalancing_Final_061719_Pages-1.pdf}.
\bibitem[{{MobilityData}(2023)}]{MobData2023}
\bibinfo{author}{{MobilityData}}, \bibinfo{year}{2023}.
\newblock \bibinfo{title}{{Mobility Database}}.
\newblock \bibinfo{note}{\url{https://database.mobilitydata.org/}}.
\bibitem[{Mohring(1972)}]{mohringOptimizationScaleEconomies1972}
\bibinfo{author}{Mohring, H.}, \bibinfo{year}{1972}.
\newblock \bibinfo{title}{Optimization and scale economies in urban bus transportation}.
\newblock \bibinfo{journal}{American Economic Review} \bibinfo{volume}{62}, \bibinfo{pages}{591--604}.
\newblock \DOIprefix\doi{10.2307/1806101}.
\bibitem[{Moloney(2021)}]{moloneyFinalBronxBus2021}
\bibinfo{author}{Moloney, S.}, \bibinfo{year}{2021}.
\newblock \bibinfo{title}{Final {{Bronx Bus Redesign Plan May Involve}} the {{Removal}} of 18 {{Percent}} of {{Bronx Bus Stops}}}.
\newblock \bibinfo{journal}{Norwood News} \URLprefix \url{https://www.norwoodnews.org/final-bronx-bus-redesign-plan-may-involve-the-removal-of-18-percent-of-bronx-bus-stops}.
\bibitem[{Morency et~al.(2011)Morency, Tr{\'e}panier~Martin and Demers}]{morencyWalkingTransitUnexpected2011}
\bibinfo{author}{Morency, C.}, \bibinfo{author}{Tr{\'e}panier~Martin, M.}, \bibinfo{author}{Demers, M.}, \bibinfo{year}{2011}.
\newblock \bibinfo{title}{Walking to transit: {{An}} unexpected source of physical activity}.
\newblock \bibinfo{journal}{Transport Policy} \bibinfo{volume}{18}, \bibinfo{pages}{800--806}.
\newblock \DOIprefix\doi{10.1016/j.tranpol.2011.03.010}.
\bibitem[{{MTA}(2022)}]{queens2022}
\bibinfo{author}{{MTA}}, \bibinfo{year}{2022}.
\newblock \bibinfo{title}{Queens Bus Network Redesign New Draft Plan}.
\newblock \bibinfo{type}{Technical Report}.
\newblock \URLprefix \url{https://new.mta.info/queens-bus-redesign-draft-plan-hi-res}.
\bibitem[{Ouyang et~al.(2014)Ouyang, Nourbakhsh and Cassidy}]{ouyangContinuumApproximationApproach2014}
\bibinfo{author}{Ouyang, Y.}, \bibinfo{author}{Nourbakhsh, S.M.}, \bibinfo{author}{Cassidy, M.J.}, \bibinfo{year}{2014}.
\newblock \bibinfo{title}{Continuum approximation approach to bus network design under spatially heterogeneous demand}.
\newblock \bibinfo{journal}{Transportation Research Part B: Methodological} \bibinfo{volume}{68}, \bibinfo{pages}{333--344}.
\newblock \DOIprefix\doi{10.1016/j.trb.2014.05.018}.
\bibitem[{Pandey et~al.(2021)Pandey, Lehe and Monzer}]{pandeyDistributionsBusStop2021}
\bibinfo{author}{Pandey, A.}, \bibinfo{author}{Lehe, L.}, \bibinfo{author}{Monzer, D.}, \bibinfo{year}{2021}.
\newblock \bibinfo{title}{Distributions of {{Bus Stop Spacings}} in the {{United States}}}.
\newblock \bibinfo{journal}{Findings} \DOIprefix\doi{10.32866/001c.27373}.
\bibitem[{Pereira et~al.(2022)Pereira, Andrade and Vieira}]{Pereira2022}
\bibinfo{author}{Pereira, R.H.}, \bibinfo{author}{Andrade, P.R.}, \bibinfo{author}{Vieira, J.P.B.}, \bibinfo{year}{2022}.
\newblock \bibinfo{title}{Exploring the time geography of public transport networks with the gtfs2gps package}.
\newblock \bibinfo{journal}{Journal of Geographical Systems} , \bibinfo{pages}{1--14}\DOIprefix\doi{10.1007/s10109-022-00400-x}.
\bibitem[{Reilly(1997)}]{reillyTransitServiceDesign1997}
\bibinfo{author}{Reilly, J.M.}, \bibinfo{year}{1997}.
\newblock \bibinfo{title}{Transit {{Service Design}} and {{Operation Practices}} in {{Western European Countries}}}.
\newblock \bibinfo{journal}{Transportation Research Record: Journal of the Transportation Research Board} \bibinfo{volume}{1604}, \bibinfo{pages}{3--8}.
\newblock \DOIprefix\doi{10.3141/1604-01}.
\bibitem[{RTD(2021)}]{RTD}
\bibinfo{author}{RTD}, \bibinfo{year}{2021}.
\newblock \bibinfo{title}{{Bus Stop Consolidation}}.
\newblock \bibinfo{howpublished}{\url{https://www.rtd-denver.com/projects/bus-stop-consolidation}}.
\bibitem[{Stewart and {El-Geneidy}(2016)}]{stewartDonStopJust2016}
\bibinfo{author}{Stewart, C.}, \bibinfo{author}{{El-Geneidy}, A.}, \bibinfo{year}{2016}.
\newblock \bibinfo{title}{Don't stop just yet! {{A}} simple, effective, and socially responsible approach to bus-stop consolidation}.
\newblock \bibinfo{journal}{Public Transport} \bibinfo{volume}{8}, \bibinfo{pages}{1--23}.
\newblock \DOIprefix\doi{10.1007/s12469-015-0112-9}.
\bibitem[{Tirachini(2014)}]{tirachiniEconomicsEngineeringBus2014}
\bibinfo{author}{Tirachini, A.}, \bibinfo{year}{2014}.
\newblock \bibinfo{title}{The economics and engineering of bus stops: {{Spacing}}, design and congestion}.
\newblock \bibinfo{journal}{Transportation Research Part A: Policy and Practice} \bibinfo{volume}{59}, \bibinfo{pages}{37--57}.
\newblock \DOIprefix\doi{10.1016/j.tra.2013.10.010}.
\bibitem[{{Transportation Research Board}(2001)}]{NAP10110}
\bibinfo{author}{{Transportation Research Board}}, \bibinfo{year}{2001}.
\newblock \bibinfo{title}{{Making Transit Work: Insight from Western Europe, Canada, and the United States -- Special Report 257}}.
\newblock \bibinfo{publisher}{The National Academies Press}, \bibinfo{address}{Washington, DC}.
\newblock \DOIprefix\doi{10.17226/10110}. \bibinfo{note}{\url{https://nap.nationalacademies.org/catalog/10110/making-transit-work-insight-from-western-europe-canada-and-the}}.
\bibitem[{Van~Nes and Bovy(2000)}]{vannesImportanceObjectivesUrban2000}
\bibinfo{author}{Van~Nes, R.}, \bibinfo{author}{Bovy, P.H.}, \bibinfo{year}{2000}.
\newblock \bibinfo{title}{Importance of objectives in urban transit-network design}.
\newblock \bibinfo{journal}{Transportation Research Record: Journal of the Transportation} , \bibinfo{pages}{25--34}\DOIprefix\doi{10.3141/1735-04}.
\bibitem[{Vuchic and Newell(1968)}]{vuchicRapidTransitInterstation1968}
\bibinfo{author}{Vuchic, V.R.}, \bibinfo{author}{Newell, G.F.}, \bibinfo{year}{1968}.
\newblock \bibinfo{title}{Rapid {{Transit Interstation Spacings}} for {{Minimum Travel Time}}}.
\newblock \bibinfo{journal}{Transportation Science} \bibinfo{volume}{2}, \bibinfo{pages}{303--339}.
\newblock \DOIprefix\doi{10.1287/trsc.2.4.303}.
\bibitem[{Wagner and Bertini(2014)}]{wagnerBenefitcostEvaluationMethod2014}
\bibinfo{author}{Wagner, Z.}, \bibinfo{author}{Bertini, R.}, \bibinfo{year}{2014}.
\newblock \bibinfo{title}{Benefit-cost evaluation method for transit stop removal}.
\newblock \bibinfo{journal}{Transportation Research Record: Journal of the Transportation} \bibinfo{volume}{2415}, \bibinfo{pages}{59--64}.
\newblock \DOIprefix\doi{10.3141/2415-06}.
\bibitem[{Wirasinghe and Ghoneim(1981)}]{wirasingheSpacingBusStopsMany1981}
\bibinfo{author}{Wirasinghe, S.C.}, \bibinfo{author}{Ghoneim, N.S.}, \bibinfo{year}{1981}.
\newblock \bibinfo{title}{Spacing of {{Bus-Stops}} for {{Many To Many Travel Demand}}.}
\newblock \bibinfo{journal}{Transportation Science} \bibinfo{volume}{15}, \bibinfo{pages}{210--221}.
\newblock \DOIprefix\doi{10.1287/trsc.15.3.210}.
\bibitem[{Wu et~al.(2022)Wu, Jin and Yang}]{Wu2022}
\bibinfo{author}{Wu, T.}, \bibinfo{author}{Jin, H.}, \bibinfo{author}{Yang, X.}, \bibinfo{year}{2022}.
\newblock \bibinfo{title}{To what extent may transit stop spacing be increased before driving away riders? referring to evidence of the 2017 nhts in the united states}.
\newblock \bibinfo{journal}{Sustainability} \bibinfo{volume}{14}.
\newblock \URLprefix \url{https://www.mdpi.com/2071-1050/14/10/6148}, \DOIprefix\doi{10.3390/su14106148}.

\end{thebibliography}
\end{document}